\documentclass[pre,twocolumn,aps,groupedaddress]{revtex4-1}
\usepackage[T1]{fontenc}
\usepackage[latin9]{inputenc}
\usepackage{array}
\usepackage{booktabs}
\usepackage{mathrsfs}
\usepackage{multirow}
\usepackage{amsmath}
\usepackage{amsthm}
\usepackage{amssymb}
\usepackage{stmaryrd}
\usepackage{graphicx}
\usepackage{latexsym}
\usepackage{textcomp}
\usepackage{mathtools}
\usepackage{xcolor}
\usepackage{xr-hyper}
\usepackage[citecolor=magenta,colorlinks=true]{hyperref}
\usepackage{lineno}
\usepackage{stackrel}
\usepackage[toc,page,titletoc]{appendix}
\usepackage{bm}

\newcommand{\lyxmathsym}[1]{\ifmmode\begingroup\def\b@ld{bold}
  \text{\ifx\math@version\b@ld\bfseries\fi#1}\endgroup\else#1\fi}

\providecommand{\tabularnewline}{\\}

\renewcommand\[{\begin{equation}} 
\renewcommand\]{\end{equation}}

\begin{document}
\title{Light and heavy particles on a fluctuating surface:\\
Bunchwise balance, irreducible sequences and local density-height correlations}

\author{Samvit Mahapatra}
\email{samvit.mahapatra@gmail.com}
\affiliation{TIFR Centre for Interdisciplinary Sciences,\\ Tata Institute of Fundamental
Research, Hyderabad 500107, India}

\author{Kabir Ramola}
\email{kramola@tifrh.res.in}
\affiliation{TIFR Centre for Interdisciplinary Sciences,\\ Tata Institute of Fundamental
Research, Hyderabad 500107, India}

\author{Mustansir Barma}
\email{barma@tifrh.res.in}
\affiliation{TIFR Centre for Interdisciplinary Sciences,\\ Tata Institute of Fundamental
Research, Hyderabad 500107, India}

\begin{abstract}

We study the early time and coarsening dynamics in the Light-Heavy model, a system consisting of two species of particles (\emph{light} and \emph{heavy}) coupled to a fluctuating surface (described by tilt fields). The dynamics of particles and tilts are coupled through local update rules, and are known to lead to different ordered and disordered steady state phases depending on the microscopic rates. We introduce a generalized balance mechanism in non-equilibrium systems, namely {\it bunchwise balance}, in which incoming and outgoing transition currents are balanced between groups of configurations. This allows us to exactly determine the steady state in a subspace of the phase diagram of this model. We introduce the concept of {\it irreducible sequences} of interfaces and bends in this model. These sequences are non-local, and we show that they provide a coarsening length scale in the ordered phases at late times. Finally, we propose a {\it local} correlation function ($\mathcal{S}$) that has a direct relation to the number of irreducible sequences, and is able to distinguish between several phases of this system through its coarsening properties. Starting from a totally disordered initial configuration, $\mathcal{S}$ displays an initial linear rise and a broad maximum. As the system evolves towards the ordered steady states, $\mathcal{S}$ further exhibits power law decays at late times that encode coarsening properties of the approach to the ordered phases. Focusing on early time dynamics, we posit coupled mean-field evolution equations governing the particles and tilts, which at short times are well approximated by a set of linearized equations, which we solve analytically. Beyond a timescale set by an ultraviolet (lattice) cutoff and preceding the onset of coarsening, our linearized theory predicts the existence of an intermediate diffusive (power-law) stretch, which we also find in simulations of the ordered regime of the system.

\end{abstract}
\date{\today}
\maketitle

\section{Introduction}
\label{introduction_section}

Phase separation, coarsening and dynamical arrest in interacting non-equilibrium systems arises in a variety of contexts in physics and biology. Examples include turbulent mixtures \cite{berti2005turbulence}, driven granular materials \cite{jaeger1996granular}, constrained systems at low temperatures \cite{bouchaud1998out,testard2014intermittent}, as well as soft matter and biological systems \cite{Tanaka2017PhaseSep}. hard-core particle systems that model several types of materials often display glassy dynamics and provide examples of unusually slow coarsening toward phase separation, however, they remain hard to characterize theoretically. 
Simple models of confined hard particles also display other non-trivial behavior such as anomalous transport properties \cite{klages2008anomalous}. For instance, driven hard-core particles in one dimension serve as useful models for transport along channels and surfaces, and have a long history of study \cite{lebowitz1988microscopic}. Their coarse grained properties have been related to the Kardar-Parisi-Zhang (KPZ) and Burgers equations that also describe the hydrodynamics of surfaces and compressible fluids \cite{MedinaZhang1989Burgers-Interface}. 

\subsection*{Fluctuating Local Drives: hard-core particles on a Fluctuating Surface} 

While globally driven systems have been well-studied, interesting effects arise in systems with a fluctuating {\it local} drive, which have been of considerable recent interest in connection with active particle systems \cite{ramaswamy2017active}. These systems display new non-equilibrium properties such as motility induced phase separation (MIPS) \cite{cates2015motility} and even complete dynamical arrest \cite{merrigan2020arrested}. However, several key questions about the dynamics of such systems remain open. In this context, one-dimensional models of particles coupled to a fluctuating surface have proved to be useful tools to study and characterize the behavior of locally driven non-equilibrium systems \cite{LR1997Sedimentation,DrosselKardar2000PhaseOrdering,Das2000fdpoPRL}. Even though many such models are built using simple local rules, an exact determination of their non-equilibrium steady states has been hard, with only a few known cases. In this paper, we introduce new theoretical approaches to study the out-of-equilibrium behavior of the Light-Heavy (LH) model, a simple lattice model consisting of two species of hard-core particles interacting with a fluctuating surface.
This model (discussed in detail below) is known to exhibit several non-equilibrium steady state phases \cite{LR1997Sedimentation,LBR2000SPS,DasBasu2001Dynamic-Scaling,Chakraborty2016FastDynamics-LH,Chakraborty2017LH-statics,Chakraborty2017LH-dynamics,Chakraborty2019CoupledModes-LH}. In particular, we present new methods which allow us to exactly determine the non-equilibrium steady state within a subspace of parameters of this model, and to derive new results on its early and late time dynamics.

Theoretically, the LH model is interesting as it has characteristics of a class of multi-species non-equilibrium systems. Moreover, the sorts of cooperative effects exhibited by the LH model, namely propagating waves and clustering reminiscent of phase separation, arise in several physical settings, as discussed below. For instance, waves and separately clustering, have been observed on the membrane of a cell as a result of the coupling between the membrane and the actin cytoskeleton \cite{PelegGov2011CellMemb-Waves}. A continuum model which couples the membrane and protein dynamics has been used to explain the formation of protrusions and protein segregation on the membrane \cite{VekslerGov2007CellMemb-Transitions}. The effect of active inclusions in an interface was studied in \cite{CagnettaEvans2018ActiveGrowth},  motivated by the formation of organized dynamical structures within the membrane of crawling cells. If particles on the membrane are passive and do not act back on the actomyosin, the result is a clustered state of a different sort with signatures of fluctuation-dominated phase ordering (FDPO) \cite{DasPolleyMadan2016Active-Medium}, in which long-range order coexists with macroscopic fluctuations. Propagating waves and clustering are also found in a system of active pumps with a two-way interaction with a fluid membrane \cite{SriramTonerProst2000Active-Membranes}. A very different context is provided by the problem of a colloidal crystal which is sedimenting through a viscous fluid. The problem involves the interplay between two coupled fields, namely the overall particle density, and the tilt field which governs the local direction of settling, with respect to gravity. A study of a lattice model of the system shows that the system displays macroscopic phase separation in steady state \cite{LR1997Sedimentation,LBR2000SPS}.

The local rules of the LH model were first defined in \cite{LBR2000SPS}. It was initially studied within a restricted subspace in \cite{LBR2000SPS,DasBasu2001Dynamic-Scaling}, and was later studied more generally in \cite{Chakraborty2016FastDynamics-LH, Chakraborty2017LH-statics, Chakraborty2017LH-dynamics, Chakraborty2019CoupledModes-LH}. The model describes light ($L$) and heavy ($H$) particles interacting with the local slopes (tilts) of a fluctuating surface. The particle-surface coupling arises as follows: surface slopes provide a dynamically evolving bias which guides particle movement, while the back-action of $L$ and $H$ particles on the surface affects its evolution. In spite of its simplicity, the system exhibits a rich phase diagram with a disordered phase and several types of ordered phases. The disordered phase arises when the back-action of particles on the surface goes opposite to the action of surface slopes on the particles. The dynamics is then dominated by mixed-mode kinematic waves. These waves constitute a generalization of kinematic waves familiar in single-component driven systems \cite{Lighthill1955kinematic1,lighthill1955kinematic2}. A recent numerical study \cite{Chakraborty2019CoupledModes-LH} shows that the decay of these waves in one-dimension typifies new universality classes beyond that of single-component systems \cite{vanBeijeren1985excess-noise, SchmittmannZia1995driven-diffusive,Schutz2001exactly-solvable, Stinchcombe2001stochastic-noneq}, with distinct dynamic exponents and scaling functions \cite{vanBeijeren2012anomalous-transport, MendlSpohn2013nlf-hydrodynamics,Spohn2014AnharmonicChains,Popkov2014SuperdiffusiveModes, Popkov2015Fibonacci-universality,Sasamoto2018Two-Species}. 

On the other hand, when the back-action of the particles on the surface is in consonance with the action of surface slopes on the particles, there is a tendency to form large $H$ particle clusters residing on large sloping segments. At late times, these segments show coarsening behavior, ultimately resulting in the formation of one of three ordered phases, in all of which $H$ particles are segregated from $L$ particles. The three phases differ from each other in the macroscopic shape of the part of the surface which holds the $L$ particles. There are strong variations in the slow approaches to the respective steady states coming from differences in the dynamics of coarsening, distinct from other instances of slow dynamics, 
for example, in constrained systems in one dimension \cite{Spohn1989Stretched-Exponential,Grynberg2001OneD-Interfaces}. Finally, the transition boundary between disordered and ordered phases exhibits FDPO \cite{Das2000fdpoPRL,Das2001fdpoPRE}.

\subsection*{Summary of Main Results}

In this paper, we use three new theoretical constructs to uncover the dynamic properties of the LH model: a balance condition which allows us to obtain exact steady states; a configuration-wise irreducible sequence whose mean length tracks growing length scales; and a local cross-correlation function which, surprisingly, is able to capture non-local features such as coarsening. The significance of our new approaches and results extends beyond the model considered in this paper and should be useful in the study of several other driven systems. Below we discuss our principal results on the evolution and coarsening properties of the density and tilt fields in the LH model.

We uncover a new subspace in the phase diagram of this model, where all configurations of the system occur with equal probability in the steady state. We show that this occurs due to a novel, general balance condition, namely \emph{bunchwise balance}, in which for every configuration, the incoming probability current from a bunch (or group) of incoming transitions is exactly balanced by the outgoing current from another uniquely specified group. This mechanism generalizes the pairwise balance mechanism that leads to the determination of the exact steady state in several driven, diffusive systems \cite{Schutz1996Pairwise} including in the Asymmetric Simple Exclusion Process (ASEP), as well as within a smaller subspace in the LH model \cite{DasBasu2001Dynamic-Scaling}. A recent submission has independently discussed the application of the bunchwise balance condition to the Zero Range Process and related systems \cite{Indranil2020Multibalance}.

We also show that the dynamics of the system can be described in terms of an {\it irreducible sequence}, a new construct in the study of driven diffusive systems, defined as follows: any configuration of the LH model can be specified by the locations and lengths of domains of $L$ and $H$ particles, and separately, of up and down tilts. More succinctly, this is encoded in a sequence which specifies the locations of walls which separate domains of $L$ and $H$ particles, and of up and down tilts. On eliminating those walls of one species (either particles or tilts) which do not enclose a wall of the other, we arrive at the irreducible sequence (IS). The IS helps to prove that bunchwise balance indeed holds for the steady state in the relevant subspace. It also helps to characterize the dynamics, as the number of elements in the IS provides a quantitative measure of coarsening during the approach to the steady state.

In the second half of the paper we focus specifically on a \emph{local cross-correlation function} $S$ defined for each configuration, as well as its counterpart $\mathcal{S}$ averaged over initial conditions, that is, surprisingly, able to capture the non-local evolution and coarsening properties of the system. This occurs because $S$ is directly related to the number of irreducible sequences. It measures the correlation between the gradient of one species (say the particles) and the density of the other (the local slope), though there is a reciprocity in the definition.  $S$ allows us to probe the evolution of the system at all time scales, ranging from very small (less than a single time step) to very large (during coarsening, as the system approaches an ordered steady state). Interestingly, the cross-correlation function has a meaning at both local and global levels. At the local level, in any configuration $S$ is non-zero at those locations of the system where dynamic evolution is possible. At the global level, the averaged correlation $\mathcal{S}$ provides a precise measure of the degree of coarsening as the system evolves towards an ordered state. We provide numerical evidence that $\mathcal{S}$ is able to distinguish and characterize the several non-trivial phases that occur in the LH model, through its coarsening properties.

We derive the exact early time evolution of $\mathcal{S}$ for totally disordered initial conditions, and determine the slope which describes the early time linear growth. We use the result to show that the subspace determined from the bunchwise balance condition in fact exhausts the set of all states with equal weights for all configurations.
We next discuss a mean field theory for the evolution of the density and tilt fields. It describes the evolution of $\mathcal{S}$ at early times surprisingly well, and reproduces the exact slopes at early times. At later times, this approximate theory departs from numerical results, reflecting a build-up of correlations that are ignored in the mean field treatment. 

This paper is organized as follows. In Section \ref{sec:The-Light-Heavy-Model} we introduce the well-studied Light-Heavy (LH) model and discuss known results as well as different regions in the phase diagram in detail. In Section \ref{bunchwise_section} we describe the bunchwise balance mechanism that produces an equiprobable steady state in the LH model. In Section \ref{S_section} we introduce the local cross-correlation function $\mathcal{S}$ and describe its observed properties in various regions of the phase diagram. In Section \ref{exact_S_section} we derive the exact early time behavior of $\mathcal{S}$ up to linear order in time, beginning from a totally disordered initial configuration. In Section \ref{linearized_mean_field_section} we derive a mean field theory for the coupled evolution of the density and tilt fields in the system, which we use to characterize the early time behavior of $\mathcal{S}$ beyond the linear regime. In Section \ref{section_late_time_S} we study the coarsening properties of the system using $\mathcal{S}$ and provide numerical evidence that such local cross-correlations encode information about the large length scale coarsening in this system. Finally, we conclude in Section \ref{conclusions_section} and provide directions for further investigation related to the methods and results discussed in this paper.

\section{The Light-Heavy Model}
\label{sec:The-Light-Heavy-Model}

\subsection{The Model}

\begin{figure}[t!]
\includegraphics[scale=0.26]{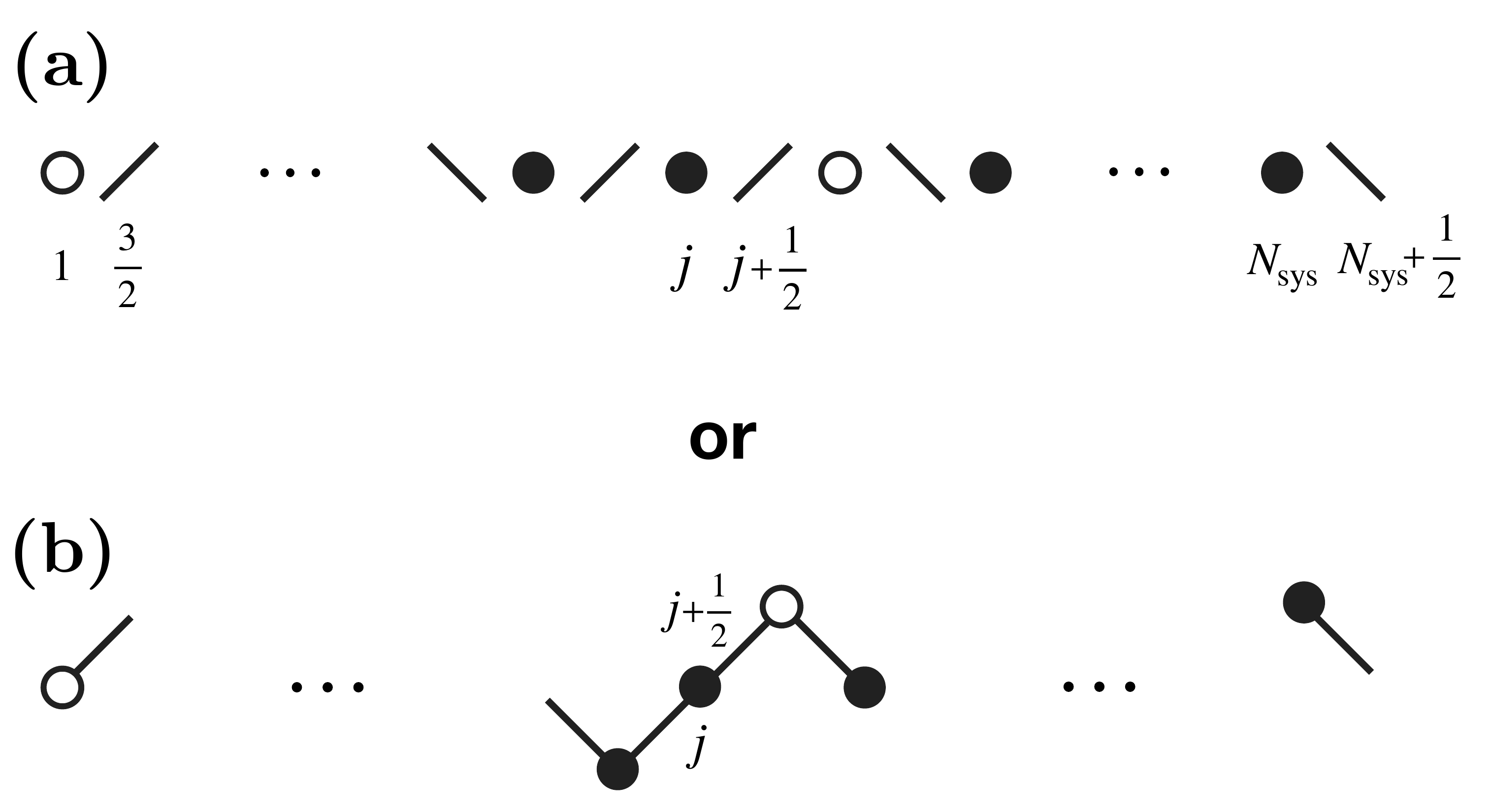}
\caption{Representations of a configuration of the LH model. {\bf (a)} The array of \emph{Light} ($\circ$) and \emph{Heavy} ($\bullet$) particles occupy integer locations $j$ of a lattice whereas \emph{Up} ($\diagup$) and \emph{Down} ($\diagdown$) tilts occupy half integer locations $j+\frac{1}{2}$. {\bf (b)} The array of tilts generates a discrete one-dimensional surface as illustrated.}
\label{Model_figure}
\end{figure}

The Light-Heavy (LH) model is a lattice system of hard-core
particles with stochastic dynamics on a one-dimensional fluctuating 
landscape \cite{Chakraborty2016FastDynamics-LH,Chakraborty2017LH-statics,Chakraborty2017LH-dynamics}.
There are two sub-lattices, with one sub-lattice containing particles represented by $\sigma$ and the other containing tilts represented by $\tau$. The particles can be of two types, light particles $\circ$ ($L$) or heavy particles $\bullet$ ($H$). The tilts can be in one of two states, an up tilt $\diagup$ or a down tilt $\diagdown$. Every configuration of tilts constitutes a discrete surface whose local slopes are given by these tilts. We label the locations of particles by integers $j$, and those of tilts by half-integers $j+\frac{1}{2}$ (see Fig.~\ref{Model_figure}).
Since there are two types of particles and two types of tilts, we may represent the state $\sigma_{j}$ of particles on each site and the state $\tau_{j+{\scriptscriptstyle \frac{1}{2}}}$ of the tilts on each site by Ising variables that take the values $\pm 1$. We assign these as follows
\begin{flalign}
\sigma_{j} & =\begin{cases}
\begin{array}{c}
-1\\
+1
\end{array} & \begin{array}{c}
Light\,\,~\circ\\
Heavy\,\,\bullet
\end{array}\end{cases} & ;\quad\tau_{j+{\scriptscriptstyle \frac{1}{2}}} & =\begin{cases}
\begin{array}{c}
-1\\
+1
\end{array} & \begin{array}{c}
Up\,\,~~~~\diagup\\
Down\,\,\diagdown
\end{array}\end{cases}
\end{flalign}

A typical LH configuration can be represented either as an array
of particles and tilts, or as particles residing on the discrete surface, as shown in Fig.~\ref{Model_figure}.
We assume that there are an even number of $N_{\text{sys}}$ sites on each sub-lattice,
with periodic boundary conditions. We consider configurations with numbers of light and heavy particles $N_{\circ}$ and $N_{\bullet}$ respectively, in a lattice size of $N_{\text{sys}}$. The numbers of \emph{$L$} and \emph{$H$} particles are conserved, but may differ. We also have $N_{\circ} + N_{\bullet} = N_{\text{sys}}$. It is convenient to define $\sigma_0 = (N_{\bullet} - N_{\circ})/N_{\text{sys}}$, such that the mean densities of light and heavy particles are given by $\rho(\circ) = \frac{1 - \sigma_0}{2}$ and $\rho(\bullet) = \frac{1 + \sigma_0}{2}$ respectively. Similarly, $N_{\diagup}$ and $N_{\diagdown}$ represent the total number of up and down tilts respectively. We further take the numbers of
up and down tilts to be equal, thus $N_{\diagup} = N_{\diagdown} = N_{\text{sys}}/2$. In this work we deal only with surfaces with no overall slope, nevertheless some of our results may also be useful for surfaces with finite overall slopes.

\begin{figure*}[t]
\includegraphics[scale=0.44]{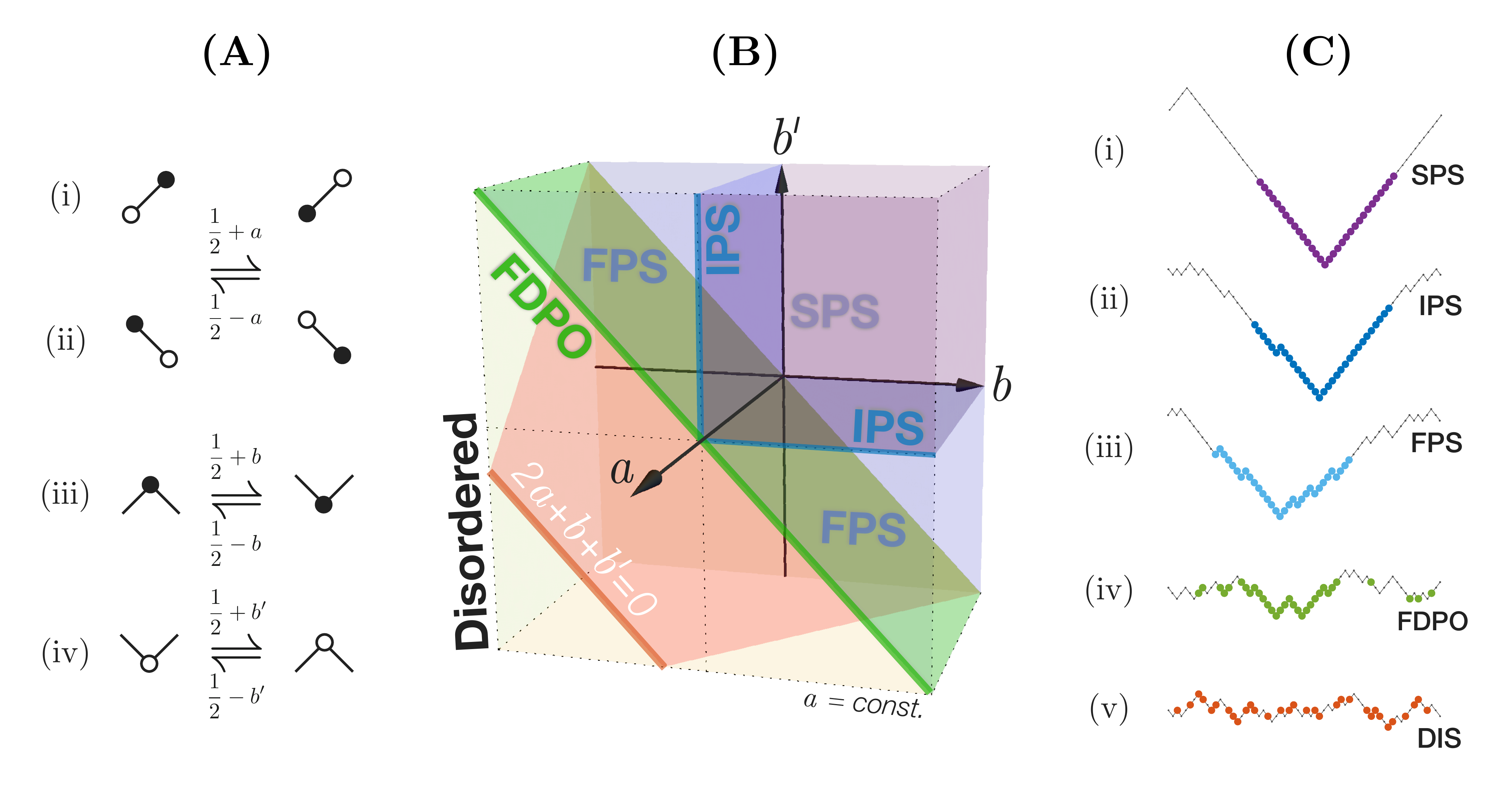}
\caption{\textbf{(A)} Update rules for the LH model. The first set of
rules (i) and (ii) involve the exchange of neighboring particles, and thus
generate particle currents. The forward and backward transitions occur with rates $\frac{1}{2}+a$ and $\frac{1}{2}-a$ respectively. The second set of rules (iii) and (iv) dictate
the interaction of particles with the surface, and correspondingly
generate tilt currents across the LH lattice. \textbf{(B)} The three-dimensional phase diagram of the LH model in the space of update parameters $a$, $b$, $b'$. The bunchwise balance condition $2 a + b +b' = 0$ represents a plane (shown in red) that lies in the disordered phase (light yellow) of the phase diagram. Although the parameter $a$ ranges from $0$ to $1/2$, the section for an intermediate value of $a=0.3$ is depicted as the exterior face in the figure. \textbf{(C)} Schematic plots of the steady state profiles with \emph{H} particles shown in color, for different phases
of the LH model. Profiles (i), (ii) and (iii) are the ordered
phases SPS, IPS and FPS respectively, with differing extents of phase
separation between the tilts, while the particles are always perfectly
phase separated. The order-disorder phase boundary in (iv) exhibits Fluctuation Dominated Phase Ordering (FDPO). long-range order prevails
in FDPO, in contrast to the disordered regime (v). }
\label{fig: phase_diagram}
\end{figure*}

The update rules for the model couple the dynamics of particle and
tilt sub-lattices [Fig.~\ref{fig: phase_diagram} (A)]. They allow
exchanges between neighboring $L$ and $H$ particles. Likewise,
exchanges between neighboring up and down tilts are also allowed,
which take a local hill to a valley, and vice versa. The update rules
involve transition probabilities controlled by the parameters $a,b$
and $b'$, which introduce biases between the forward and reverse
updates. The parameter $a$ modulates the exchanges between neighboring
$L$ and $H$ particles [rules (i) and (ii) in Fig.~\ref{fig: phase_diagram} (A)] depending on the sign of the tilt between them. In a microscopic time step $[t, t + \Delta t]$, the forward transitions in rules (i) and (ii) occur with a probability $\left( \frac{1}{2}+a \right) \Delta t$, whereas the reverse transitions occur with a probability $\left( \frac{1}{2}-a \right) \Delta t$. The parameters $b$ and $b'$ modulate exchanges between neighboring up and down tilts [rules (iii) and (iv) in Fig.~\ref{fig: phase_diagram} (A)], depending on whether $L$ or $H$ particles occur in between. Alternatively, the tilt exchange rates embody the propensities of the particles to
push the surface upwards or downwards, according to the signs of $b,\,b'$. 

It is apparent that reversing the signs of $a$, $b$, $b'$ is the
same as interchanging the up and down tilts. In this paper, we assume
$a>0$, which implies that the \emph{$H$} particles tend to slide
downwards by displacing the $L$ particles upwards. The values of
$b'$ (and $b$) can be positive, negative, or zero, and thus the
$L$ (or $H$) particles may push the surface either upwards for $b'>0$
(or $b<0$), downwards for $b'<0$ (or $b>0$), or to an equal extent for $b=-b'$, or not at all for $b'=0$
(or $b=0$). 

Several earlier studies of models in different physical contexts are included in the LH model as special cases. First, this model was defined in the context of sedimenting colloidal crystals \cite{LBR2000SPS}, in which case the two species represent gradients of the longitudinal and shear strains. However, only the case $b = b' > 0$ was studied in \cite{LBR2000SPS} and shown to give rise to an exceptionally robust sort of phase separation.  Secondly, when $b+b' = 0$, the surface is pushed to an equal extent by \emph{L} and \emph{H} particles irrespective of their placements. Thus the problem reduces to that of \emph{H} particles sliding down a fluctuating interface, which itself evolves autonomously according to KPZ dynamics (if $b'=-b$ is nonzero) or Edwards-Wilkinson dynamics (if $b'= -b = 0$) \cite{Das2000fdpoPRL,Das2001fdpoPRE,kapri2016op-scaling}. Thirdly, the problem of a single active slider on a fluctuating surface was studied in \cite{CagnettaEvans2019Single-Slider} as a model of membrane proteins that activate cytoskeletal growth. This model corresponds to the case of a single \emph{H} particle in the LH model.

\subsection{Phase Diagram}

Figure ~\ref{fig: phase_diagram}~(B) shows that within the 3-dimensional parameter
space ($a$, $b$, $b'$), the system in steady state exhibits several
ordered and disordered phases. In all the ordered phases SPS, IPS
and FPS [profiles (i)-(iii) in Fig.~\ref{fig: phase_diagram} (C)], the
$L$ and $H$ particles are completely phase separated. 
These constitute pure phases, as typically only $L$ or $H$ particles are present in the bulk of each phase.
However, the
extent to which the up and down tilts are ordered differs from one
phase to another, leading to different forms of the height profile of the surface at the macroscopic level.  
Further, the phase separated regions in the tilt
profile behave as reservoirs, generating tilt currents whose uniform
flow throughout the system causes the steady state landscape to drift
downwards collectively. The drift velocities depend on the magnitudes
of tilt current for the different ordered phases. Below, we briefly
summarize the general features of the different steady state phases,
namely the nature of their phase separation and overall movement of
their landscapes. 

\emph{Strong Phase Separation} \textbf{(SPS)}, ($b+b'>0$, $b$ and $b'>0$):
In this phase, the $L$ particles push the surface upwards, while
the $H$ particles push the surface downwards \cite{LR1997Sedimentation,LBR2000SPS,Chakraborty2017LH-statics}. Like the particles,
the tilts also cleanly phase separate into two co-existing ordered
regions of up and down tilts [profile (i) in Fig.~\ref{fig: phase_diagram}~(C)].
The steady state configurations comprise a single macroscopic tilt
valley with \emph{$H$} particles, and a macroscopic hill with \emph{$L$}
particles. The complete state is thus near-perfectly ordered, except
at the interfaces between ordered regions. The system is essentially static
in steady state, apart from an exponentially small tilt current $\sim\exp(-\lambda\,N_{\text{sys}})$,
where $\lambda$ is a constant which depends on the update parameters. The relaxation to the steady state in this regime is logarithmically slow \cite{LBR2000SPS,shaurithesis2020}.

\emph{Infinitesimal current with Phase Separation} \textbf{(IPS)}, ($b+b'>0$,
$b$ or $b'=0$): Only one species, i.e. either $L$ or $H$ particles push the surface upward or downward respectively \cite{Chakraborty2016FastDynamics-LH,Chakraborty2017LH-statics,Chakraborty2017LH-dynamics}. The tilts are separated into three
co-existing regions; two of these regions are perfectly ordered, whereas
the third region is disordered. For the case $b'=0$, the two ordered
regions reside within the $H$ cluster, forming a single macroscopic
valley, while the disordered region spans the $L$ cluster [profile
(ii) in Fig.~\ref{fig: phase_diagram} (C)]. The behavior of the disordered
tilt region can be mapped to an open chain Symmetric Exclusion Process
(SEP) \cite{Derrida2002openSEP} with up and down tilts always fixed at the two ends. This is
seen easily by identifying the up tilts with particles and down tilts
with holes in the SEP. Consequently, the up tilt density in the disordered
region varies linearly with a gradient $\sim1/N_{\text{sys}}$, giving rise
to an `infinitesimal' tilt current which scales as $\sim1/N_{\text{sys}}$.
Since the mean current is uniform in the steady state, this translates
to a downward drift of the system with an average velocity $\sim1/N_{\text{sys}}$. For the case $b=0$, $b'>0$, the same features follow, except that the $L$
and $H$ particles are interchanged and the system drifts upward.

\emph{Finite current with Phase Separation} \textbf{(FPS)}, ($b+b'>0$, $b$ or $b'<0$): Both \emph{$L$} and \emph{$H$} particles push the surface
in the same direction, but to unequal extents \cite{Chakraborty2016FastDynamics-LH,Chakraborty2017LH-statics,Chakraborty2017LH-dynamics}. The tilt profile is
again made up of three distinct regions. However, in contrast to IPS,
two of these phases are imperfectly ordered, i.e. are not pure phases, while the third phase
is disordered. For the case $b'<0$, the two ordered phases form a
valley beneath the $H$ particles with smaller slopes, because of
imperfect ordering. The disordered phase resides along with the $L$
cluster [profile (iii) in Fig.~\ref{fig: phase_diagram}~(C)], and maps
to an Asymmetric Simple Exclusion Process (ASEP) in the maximal current
phase \cite{Derrida1992openASEP}. This induces a uniform tilt current
that remains finite in the thermodynamic limit, and depends on the
update parameters. The finite current causes the system to drift downwards
with a constant velocity. Other than $L$ and $H$ particles having
interchanged, the same features follow for the case $b<0$.

\emph{Fluctuation Dominated Phase Ordering} \textbf{(FDPO)}, ($b+b'=0$): This
is the order-disorder separatrix, and shows fluctuation dominated phase ordering [profile (iv) in Fig.~\ref{fig: phase_diagram}~(C)]. In this regime, both \emph{L}
and \emph{H} particles push the surface to an equal extent, and thereby
cause the surface to evolve as if it is autonomous with KPZ dynamics \cite{Chakraborty2016FastDynamics-LH}. Thus, the surface is completely disordered, and drifts
downwards with a finite rate. The particles behave as passive scalars,
directed by the dynamics of the fluctuating surface. This results in a statistical state which exhibits FDPO, in which long-range order coexists with extremely large fluctuations, leading to a characteristic cusp in the scaled two-point correlation function. The largest particle clusters are macroscopic with typical size of order $\sim N_{\text{sys}}$. However, these macroscopic clusters reorganise continuously in time. The properties of FDPO have been studied in detail in \cite{Das2000fdpoPRL,Das2001fdpoPRE,ManojG2003KPZ-2D,chatterjee2006passive-sliders,kapri2016op-scaling}. It has been invoked in the study of a variety of equilibrium and non-equilibrium systems, including active nematics and vibrated rods \cite{DasRajesh2012GiantNFluctuations}, actin-stirred membranes \cite{DasPolleyMadan2016Active-Medium}, inelastically colliding particles \cite{DasRajesh2007GranularGas} as well as Ising models with long-range interactions \cite{Barma2019Mixed-Order}.

\emph{Disordered Phase}, ($b+b'<0$): In the disordered phases, the surface-pushing
tendencies of the $L$ and $H$ particles oppose the tendency of the
$H$ particles to drift downwards \cite{DasBasu2001Dynamic-Scaling}. This does not allow large, ordered
structures to form throughout the system, and both particle and tilt
profiles are completely disordered; the steady state is characterised
by the absence of long-range correlations [profile (v) in Fig.~\ref{fig: phase_diagram} (C)].
Kinematic waves propagate across the particle and tilt sub-lattices
in steady state \cite{DasBasu2001Dynamic-Scaling}, and the nature of their decay changes across the disordered phase \cite{Chakraborty2019CoupledModes-LH}, giving rise to several new dynamical universality classes \cite{vanBeijeren2012anomalous-transport, MendlSpohn2013nlf-hydrodynamics,Spohn2014AnharmonicChains,Popkov2014SuperdiffusiveModes, Popkov2015Fibonacci-universality,Sasamoto2018Two-Species}.

It has recently been hypothesized \cite{Barma2019Mixed-Order} that the phase transition from the disordered phase to an ordered phase in the LH model, across the FDPO transition locus, is a mixed order transition, which shows a discontinuity of the order parameter as well as a divergent correlation length at the transition. Such a mixed order transition occurs across the FDPO locus in a 1D Ising model with short and truncated long-ranged interactions \cite{Barma2019Mixed-Order}. In the LH model, recall that all the ordered phases, FPS, IPS and SPS, display complete phase segregation of $L$ and $H$ particles, implying maximum order. In the disordered phase, there is no $L$-$H$ segregation on large scales, implying that the order parameter vanishes. Thus the order parameter shows a strong discontinuity across the transition. While the divergence of the correlation length remains to be established, the trends displayed by a local cross-correlation function reported below (Sections \ref{S_section} and \ref{linearized_mean_field_section}) are consistent with this scenario.

\subsubsection*{Subspaces with exactly known steady states}

There are two special subspaces within the phase diagram, where the
steady state of the system can be determined exactly. 
The first subspace resides within the SPS phase. Here the system satisfies the condition of detailed balance. The steady state measure is $\sim\exp\left(-\beta\mathcal{H}\right)$,
for a Hamiltonian $\mathcal{H}$ defined with potential energies assigned
to each \emph{H} particle,  in proportion to their vertical heights. Detailed balance in SPS was first established in \cite{LBR2000SPS} on the locus $b=b'$ which ensures
interchange symmetry between $\sigma$ and $\tau$. This was later generalized to the case of variable
particle densities in \cite{Chakraborty2017LH-statics}. In this paper we show there is a second subspace in the disordered phase which is given by the condition  $R =2 a + b + b'=0$. 
In Section \ref{bunchwise_section},
we establish equiprobability measure in this subspace using a new condition known as bunchwise balance, substantially enlarging the earlier result which was proved on the locus $b = b' = -a$ \cite{DasBasu2001Dynamic-Scaling}.


\section{Bunchwise Balance}
\label{bunchwise_section}

In this Section we discuss a region in the phase diagram of the LH model in which the steady state is described by an equiprobable measure over all accessible configurations of the system. We show that this occurs due to a novel mechanism in which the incoming probability currents into a given configuration from a {\it group} of configurations, are balanced by outgoing transition currents to {\it another group} of configurations. We term
this condition ``bunchwise balance'', and show that it occurs when 
\begin{equation}
R = 2 ~a + b +b' = 0.
\label{bunchwise_balance_condition}
\end{equation}
The above condition maps out a plane in the phase diagram of the LH model, as shown in Fig.~\ref{fig: phase_diagram} (B).
We prove that Eq.~(\ref{bunchwise_balance_condition})
is a sufficient condition for this balance to hold, implying that in steady state, all configurations are sampled with equal probability. In Section \ref{exact_S_section}, we also establish this is a necessary condition for equiprobability of the steady state measure over all accessible configurations.

\subsection{Steady State Balance}
We begin by analyzing the evolution equation of a general Markov process, represented as a Master equation of the form
\begin{equation}
\frac{d}{dt}  p (c,t)  = \sum_{c'} M(c,c')  p (c',t).
\label{Eq_markov_evolution2}
\end{equation}
The elements of the evolution matrix are given by the microscopic transition rates between configurations, for example the rates of particle and tilt updates described in Section II.
The matrix elements are then explicitly given by
\begin{eqnarray}
\nonumber
M(c,c') &=& ~~r_{c \leftarrow c'} ~~~~~~~~~~~~\textmd{for}~~ c \neq c',\\
M(c,c) &=& -\sum_{c'} r_{c \rightarrow c'}.
 \label{Eq_markov_elements}
\end{eqnarray}
where $ r_{c \rightarrow c'}$ is the transition rate from configuration $c$ to configuration $c'$. Here $r_{c \rightarrow c'}$ and $r_{c' \leftarrow c}$ represent the same transition rate.
The outgoing probability current from configuration  $c \to c'$ and the incoming current from $c \leftarrow c''$  are given by
\begin{eqnarray}
\nonumber
j_{c\rightarrow c'} &=& p(c,t) ~r_{c\rightarrow c'},\\
j_{c\leftarrow c''} &=& p(c'',t) ~r_{c\leftarrow c''}.
\label{Eq_forward_backward_current}
\end{eqnarray}

\begin{figure}[t!]
\hspace{1cm}
\includegraphics[scale=0.2]{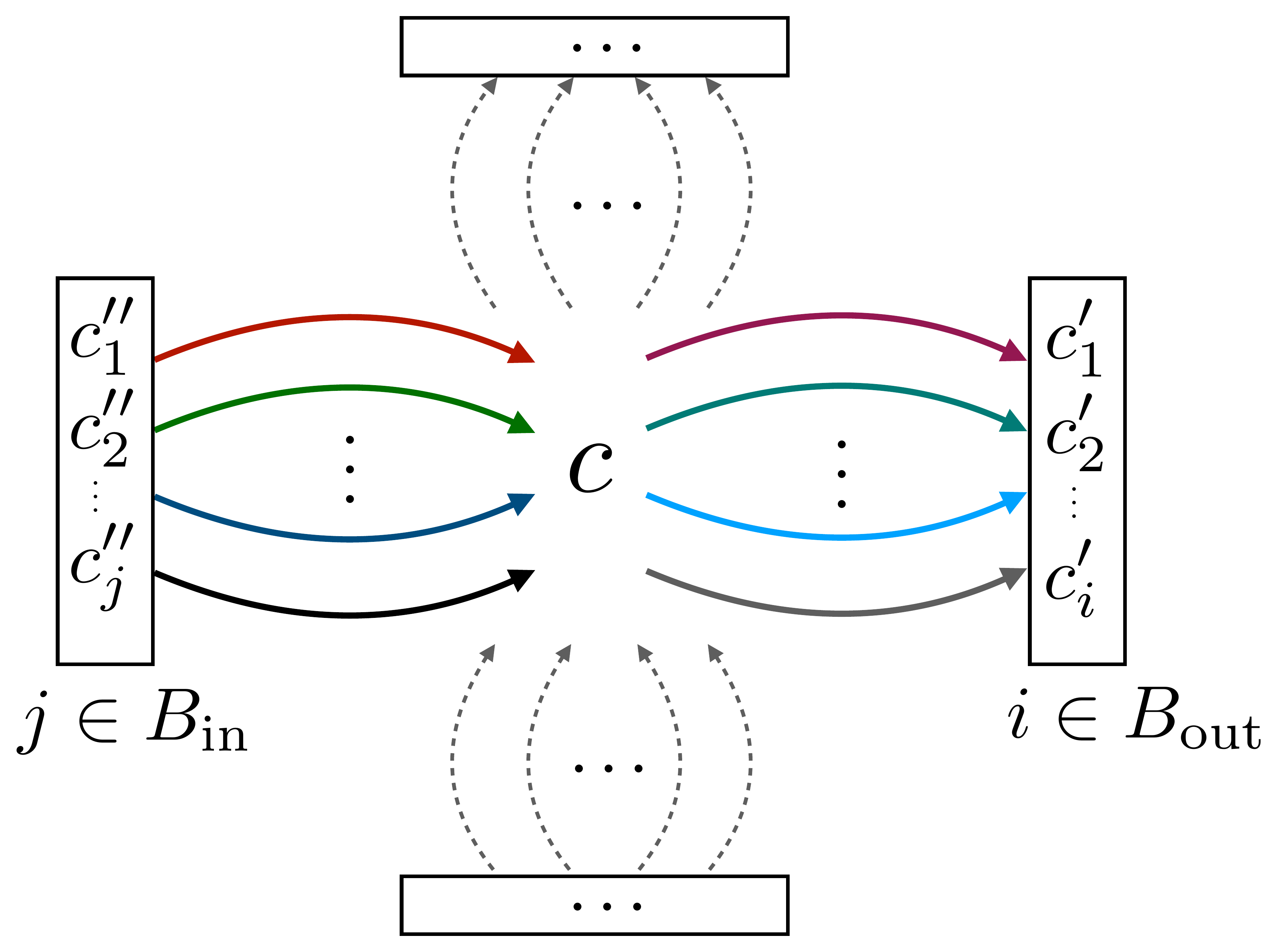}
\caption{A schematic illustration of the bunchwise balance mechanism. For every configuration $c$, the incoming probability currents from a group of configurations $\{c''_j\}$ with $j \in B_{\text{in}}$ (the incoming bunch), are balanced by outgoing currents to another uniquely identified group of configurations $\{c'_i\}$ with $i \in B_{\text{out}}$ (the outgoing bunch).
}
\label{Fig_Bunchwise_Explanatory2}
\end{figure}

The net incoming and outgoing probability currents from every configuration $c$ per unit time are then given by
\begin{eqnarray}
\nonumber
 j_{\textmd{out}}(c,t) &=& \sum_{c'}p(c,t) r_{c \rightarrow c'},\\
 j_{\textmd{in}}(c,t) &=&\sum_{c''}p(c'',t) r_{c \leftarrow c''}.
\end{eqnarray}
 In steady
state, the net incoming and outgoing probability currents at any time are equal for every configuration
and hence the probability of occurrence of a configuration $p(c,t)$ becomes independent of time, i.e. $j_{\textmd{in}}(c,t) = j_{\textmd{out}}(c,t)$. Therefore in steady state
\begin{equation}
 \sum_{c'}p(c',t) r_{c \rightarrow c'} = \sum_{c''}p(c'',t) r_{c \leftarrow c''}.
\label{Eq_steady_state_current_balance}
\end{equation}
We refer to this condition as {\it steady state balance}, as all transitions in and out of any configuration $c$ need to be summed over in order for the above balance condition to hold.

\begin{figure}[t!]
\includegraphics[scale=0.2]{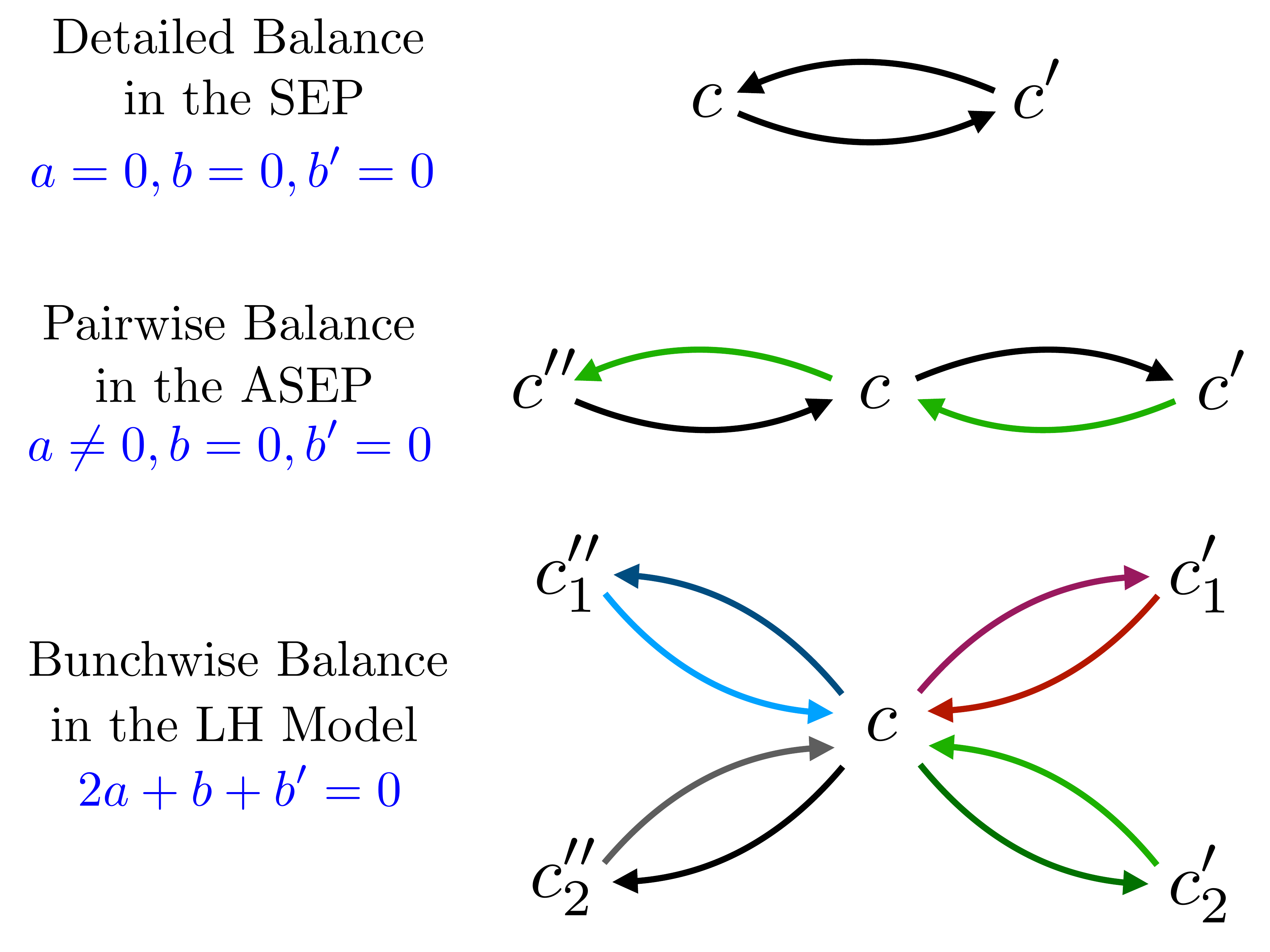}
\caption{A schematic illustration of the different balance mechanisms that are displayed by the LH model. (i) Detailed balance in the Simple Exclusion Process (SEP) in which the forward and reverse probability currents between any two configurations are equal (ii) Pairwise balance in the Asymmetric Simple Exclusion Process (ASEP) in which the incoming current from one configuration is balanced by the outgoing current to another uniquely identified configuration and (iii) Bunchwise balance in the Light-Heavy (LH) model in which incoming currents from a group of configurations, are balanced by outgoing currents to another uniquely identified group of configurations. 
}
\label{Fig_Bunchwise_Explanatory}
\end{figure}

 There are several ways in which the steady state balance condition in Eq.~(\ref{Eq_steady_state_current_balance}) can be achieved. We focus on three cases that occur in the LH model: (i) Detailed balance in which the forward and reverse probability currents between any two configurations are equal. This is given by
\begin{eqnarray}
~~~~~~j_{c \to c'} &=& j_{c \leftarrow c'}.~~~~~~~~~~~~~
{\begin{array}{c}
{\scriptstyle \mathbf{Detailed}}\\
\mathbf{{\scriptstyle Balance}}
\end{array}}
\label{eq_detailed_balance}
\end{eqnarray}
(ii) Pairwise balance in which the incoming current from one configuration is balanced by the outgoing current to another uniquely identified configuration \cite{Schutz1996Pairwise}. This is given by
\begin{eqnarray}
~~~~~~j_{c \to c'} &=& j_{c \leftarrow c''}.~~~~~~~~~~~~~
{\begin{array}{c}
{\scriptstyle \mathbf{Pairwise}}\\
\mathbf{{\scriptstyle Balance}}
\end{array}}
\label{eq_pairwise_balance}
\end{eqnarray}
Finally we introduce (iii) Bunchwise balance in which incoming currents from a {\it group} of configurations, are balanced by outgoing currents to another uniquely identified group of configurations. This mechanism is illustrated in Fig.~\ref{Fig_Bunchwise_Explanatory2} and is given by the condition
\begin{eqnarray}
\sum_{i \in B_{\text{out}}} j_{c\rightarrow c_i'} =  \sum_{j \in B_{\text{in}}} j_{c \leftarrow c_j''}. ~~~~~\mathbf{\begin{array}{c}
{\scriptstyle \mathbf{Bunchwise}}\\
\mathbf{{\scriptstyle Balance}}
\end{array}}
\label{eq_bunchwise_balance}
\end{eqnarray}
Above, the sum $i$ is over outgoing transitions belonging to a ``bunch'' $B_{\text{out}}$ whereas the sum $j$ is over incoming transitions belonging to a uniquely identified ``bunch'' $B_{\text{in}}$, such that the probabilities from these bunches are exactly equal.
Each nonzero current $j_{c\rightarrow c_i'}$ occurs in one and only one outgoing bunch, and likewise each current $j_{c \leftarrow c_j''}$  occurs once and only once in an incoming bunch. Each of the balance mechanisms discussed above appear in the LH model, and have been summarized in Fig.~\ref{Fig_Bunchwise_Explanatory}.

In some systems, in addition to the above condition, the microscopic {\it rates} also balance as
\begin{equation}
 \sum_{c'} r_{c \rightarrow c'} = \sum_{c''} r_{c \leftarrow c''}.
\label{Eq_steady_state_rate_balance}
\end{equation}
This special condition holds in a wide class of systems including the LH model in some ranges of the parameter space as we show below. It is easy to see that Eq.~(\ref{Eq_steady_state_rate_balance}) along with  Eq.~(\ref{Eq_steady_state_current_balance}) yield the time-independent solution $p(c,t) = \frac{1}{N_c}$ where $N_c$ represents the total number of configurations, i.e. all configurations occur with equal probability. Since the steady state of the Markovian dynamics governed by Eqs.~(\ref{Eq_markov_evolution2}) and (\ref{Eq_markov_elements})  is unique, at large times the system converges to an equiprobable measure over configurations.
 
\subsection{Interfaces and Bends in the LH Model}

Consider a configuration $c$ in the LH model, represented as an ordered list of occupied and unoccupied sites as well as up and down tilts on the bonds
\begin{equation}
c \equiv \hdots~\diagup~\circ~\diagdown~\circ~\diagdown~\bullet~\diagdown~\bullet~\diagdown~\bullet~\diagdown~\circ~\diagup~\hdots
\label{Eq_config_c}
\end{equation}
The evolution rules of the LH model described in Section \ref{sec:The-Light-Heavy-Model} only allow local updates at the {\it interfaces} that separate occupied and unoccupied sites, as well as at {\it bends} that separate regions of positive and negative tilts. In order to study the dynamics of the system it is therefore convenient to parametrize the configurations in the LH model in terms of variables that describe these interfaces and bends.
In order to parametrize $\sigma_j$ on the sites $j$, we introduce {\it interface} variables $\mathcal{I}_{j+\frac{1}{2}}$  on the bonds $j+\frac{1}{2}$ of the lattice. The state of the interface at each bond is determined by the density variables $\sigma_j$ and $\sigma_{j+1}$ on the adjacent sites as
\begin{equation}
 \mathcal{I}_{j+\frac{1}{2}} = (\sigma_{j+1} - \sigma_{j})/2.
\end{equation}
We note that the particle configurations are now represented by variables that can take on three values $0, \pm 1$ at every bond  in comparison with the site occupation variables which were represented by two states $-1$ or $1$. However, the interface variables satisfy constraints that ensure the one to one mapping between the two variables: namely that any interface can only be followed by a zero interface or an interface with opposite sign. The periodic boundary conditions of the system ensure that there are equal numbers of positive and negative interfaces.
For simplicity of presentation, we represent the three states of an interface as
\begin{align}
    \mathcal{I}_{j+\frac{1}{2}}   =
    \begin{cases}
        ~~1~~~~\equiv~~( \\
      ~~0~~~~\equiv~~.  \\
       -1~~~~\equiv~~ ) 
    \end{cases}
    \label{eq_interfaces}
\end{align}


Similarly, in order to parametrize the tilts $\tau_{j+\frac{1}{2}}$, we introduce {\it bend} variables $\mathcal{B}_{j}$  on the sites $j$ of the lattice. The state of the bend at each site is determined by the tilt variables $\tau_{j-\frac{1}{2}}$ and $\tau_{j+\frac{1}{2}}$ on the adjacent bonds as
\begin{equation}
 \mathcal{B}_{j} = (\tau_{j-\frac{1}{2}} - \tau_{j+\frac{1}{2}})/2.
\end{equation}
Once again these new variables can take on three values $0, \pm 1$ at every site, with a constraint that a  bend can only be followed by a zero bend or a bend with the opposite sign. The periodic boundary conditions of the system ensure that there are equal numbers of positive and negative bends.
Again, for simplicity of presentation, we represent the three states of a bend at every site as
\begin{align}
    \mathcal{B}_{j}   =
    \begin{cases}
        ~~1~~~~\equiv~~ \langle \\
      ~~0~~~~\equiv~~.  \\
       -1~~~~\equiv~~ \rangle 
    \end{cases}
    \label{eq_bends}
\end{align}
Above we have represented both interfaces and bends with values of $0$ with the same "."  symbol, since no local updates are possible at these positions.
We note that the above mapping from a configuration of densities and tilts to interfaces and bends $\{ \{ \sigma \},\{ \tau \} \} \to \{ \{ \mathcal{I} \},\{ \mathcal{B} \} \}$ is one to one and invertible.
In terms of these new interface and bend variables, the configuration in Eq.~(\ref{Eq_config_c}) simplifies to
\begin{equation}
c \equiv \hdots ~\rangle~. ~. ~. ~ ( ~.~.~.~.~.~)~\langle~\hdots
\label{Eq_config_c2}
\end{equation}
Conveniently, the only transitions in and out of this configuration $c$ occur through the updates of these  ``brackets" representing the interfaces and bends that uniquely specify $c$. The particle updates described in Section \ref{sec:The-Light-Heavy-Model},  convert the interfaces at a bond to a triad centered on the same bond as $( \rightarrow ( ) ($ and $) \rightarrow ) ( )$ respectively. Similarly, the tilt updates convert the bends at a site to a triad centered on the same site as $\langle \rightarrow \langle \rangle \langle$ and $\rangle \rightarrow \rangle \langle \rangle$ respectively. The motion of interfaces and bends through the system thus proceeds through their creation and annihilation at adjacent locations, with open and closed brackets of the same type annihilating if they reach the same position. 
We also note that the transition rates for the interfaces depend on the enclosing bends since, depending on the local slope, particles are biased to move either leftwards or rightwards. Similarly, depending on whether a site contains a light or heavy particle, the update of bends of a particular sign are favored over the other. 

\subsection{Pairwise Balance in the ASEP}

Pairwise balance is a condition found to hold in the steady state of several driven diffusive systems \cite{Schutz1996Pairwise}, including the locus $b=b'=-a$ in the LH model \cite{DasBasu2001Dynamic-Scaling}. It can be used to exactly determine the steady state of a well-known non-equilibrium 
model describing the biased motion of particles in one dimension, namely the Asymmetric Simple Exclusion Process (ASEP). The ASEP is a special case of the LH model obtained when all the tilts point in the same direction.
Below we use the ASEP as an example to illustrate the condition of pairwise balance, and also to illustrate the bracket notation that we use in analyzing the more general LH model. When all the tilts in the LH model are equal (all positive or all negative), and the particles do not influence the surface evolution (i.e. $b = b' = 0$), the dynamics of the particles in the LH model can be mapped onto the ASEP, with right and left hopping rates  $p = \frac{1}{2} + a$ and $q = \frac{1}{2} - a$ respectively. The ASEP displays an equiprobable steady state, satisfying Eq.~(\ref{Eq_steady_state_rate_balance}) in two ways as we show below: detailed balance when $p = q$ when the model reduces to the Simple Exclusion Process (SEP), and pairwise balance when $p \neq q$.

We begin with the simplest case when there is no left-right bias, i.e. $a = 0, b = 0, b' = 0$. In this case the transition rates between any two configurations $c$ and $c'$ are given by
\begin{equation}
r_{c \to c'} = r_{c \leftarrow c'} = \frac{1}{2}.
\end{equation}
These rates trivially satisfy Eq.~(\ref{Eq_steady_state_rate_balance}), implying an equiprobable steady state. This steady state solution along with the above rates then implies the detailed balance condition given in Eq.~(\ref{eq_detailed_balance}).

We next consider the case of the ASEP with a bias i.e. $a \ne 0, b = 0, b' =0$.
In this case, the forward and reverse rates between any two configurations are biased as $\frac{1}{2} \pm a$, violating detailed balance. However, the microscopic rates are still balanced as we show below. Consider a general configuration of the ASEP, described by only density variables $\{ \sigma \}$ and correspondingly only by interface variables  $\{ \mathcal{I} \}$ as
\begin{equation}
c \equiv \hdots ~( ~.~.~.~) ~. ~. ~ ( ~.~.~.~.~.~)~\hdots
\label{Eq_config_c3}
\end{equation}

\begin{figure}[t!]
\includegraphics[scale=0.2]{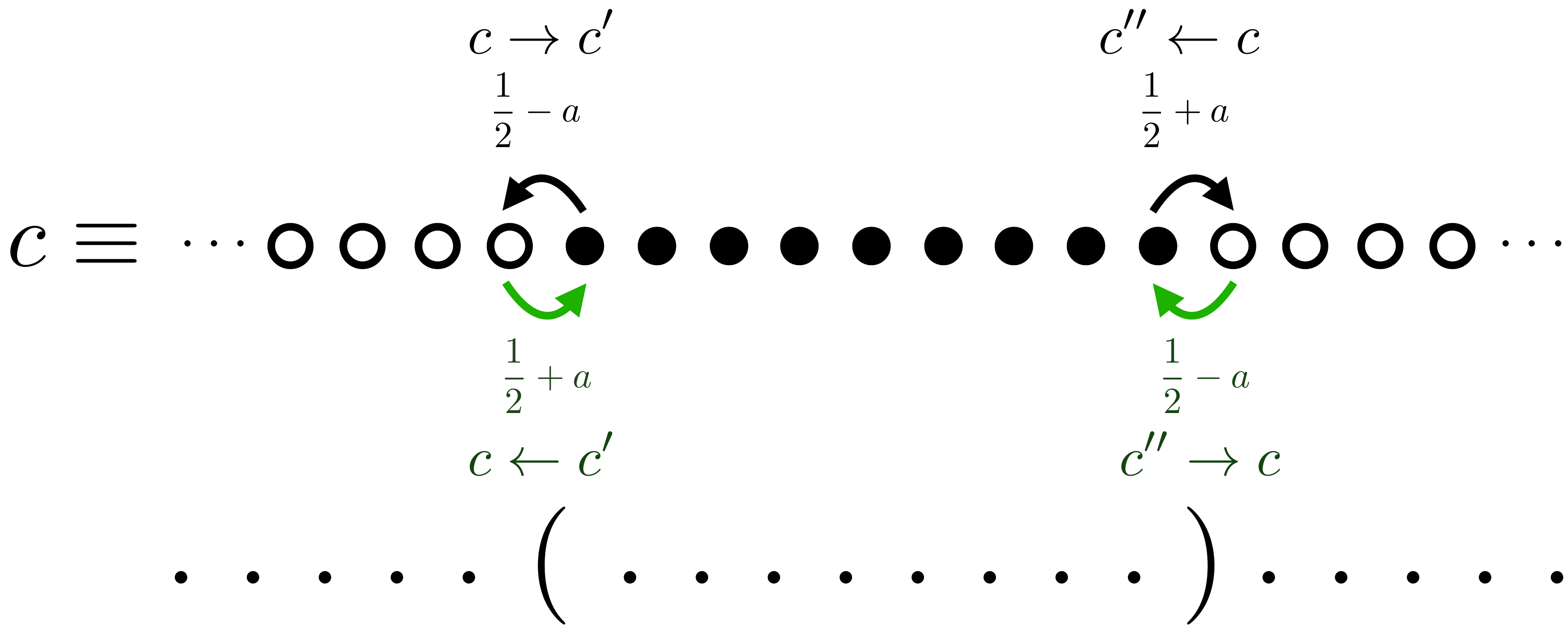}
\caption{A schematic illustration of the pairwise balance condition in the ASEP model. The black arrows represent transitions out of the configuration $c$, while the green arrows represent transitions into the configuration. In this case, each transition to a configuration $c'$ is balanced by a corresponding incoming transition from configuration $c''$ with the same rate. The configuration in terms of interfaces is represented below. Transitions between interface pairs $()$ are pairwise balanced.
}
\label{pairwise_balance_figure}
\end{figure}

To uniquely classify all the transitions of this configuration, we begin at the origin and  look at the enclosing interfaces.
Suppose moving rightwards from the origin the first bracket encountered is $($, we move forward until the next $)$ is encountered. Whereas if the first bracket encountered is $)$, we move backward until the next $($ is encountered. In this manner, every transition from this configuration, parametrized by the interfaces, has a unique corresponding pair.
An $($ interface, as shown in Fig.~\ref{pairwise_balance_figure}, represents a hole-particle interface which can be updated resulting in a configuration $c'$ with rate $r_{c \to c'} = \frac{1}{2} - a$. The reverse transition occurs with a rate $r_{c \leftarrow c'} = \frac{1}{2} + a$.
 In order to find a corresponding {\it unique} transition with the same rate, we consider the corresponding $)$ interface in the pair. This represents a particle-hole interface which can be updated, with rate $r_{c \to c''} = \frac{1}{2} + a$. Crucially, the reverse transition occurs with a rate $r_{c \leftarrow c''} = \frac{1}{2} - a$ balancing the original transition. Similarly, the reverse transitions corresponding to this pair are balanced as well. Therefore the transition rates within each such $()$ pair are balanced. 
This procedure can be extended to every interface in the configuration $c$, therefore corresponding to any outgoing transition from configuration $c \to c'$, one can identify a unique incoming transition $c \leftarrow c''$ with the same microscopic rate, i.e. 
\begin{equation}
r_{c \to c'} = r_{c \leftarrow c''}.
\end{equation}
Therefore in this case as well Eq.~(\ref{Eq_steady_state_rate_balance}) is satisfied implying an equiprobable steady state. This steady state solution along with the above rates then implies the pairwise balance condition for probability currents as given in Eq.~(\ref{eq_pairwise_balance}).

\subsection{Bunchwise Balance in the LH Model}

Finally we consider the case of the LH model with $a \ne 0, b \ne 0, b'  \ne 0$.
As opposed to the ASEP, in this case the transition rates for the interfaces are affected by the enclosing bends. For interfaces that are enclosed by (i.e. lie within consecutive bends) of the type $\langle \rangle$, the transition rates are given by
\begin{eqnarray}
\nonumber
\bm{\langle} ~~(~~\bm{\rangle} ~~\xrightleftharpoons[\frac{1}{2} -a]{\frac{1}{2} +a}~~ \bm{\langle} ~( ) (~ \bm{\rangle}  \\
\bm{\langle} ~~)~~\bm{\rangle} ~~\xrightleftharpoons[\frac{1}{2} +a]{\frac{1}{2} -a}~~ \bm{\langle}~) ( )~ \bm{\rangle} 
\label{Eq_interface_rules_1}
\end{eqnarray}
Whereas, when an interface is enclosed by $\rangle \langle$ bends, the transition rates are given by
\begin{eqnarray}
\nonumber
\bm{\rangle}~~(~~\bm{\langle} ~~\xrightleftharpoons[\frac{1}{2} +a]{\frac{1}{2} -a}~~ \bm{\rangle} ~()(~\bm{\langle}  \\
\bm{\rangle}~~)~~\bm{\langle} ~~\xrightleftharpoons[\frac{1}{2} -a]{\frac{1}{2} +a}~~ \bm{\rangle} ~)()~ \bm{\langle} 
\label{Eq_interface_rules_2}
\end{eqnarray}
Similarly,  for bends  enclosed by interfaces of the type $( )$, the transition rates are given by
\begin{eqnarray}
\nonumber
\bm{(} ~~ \langle~~\bm{)} ~~\xrightleftharpoons[\frac{1}{2} +b]{\frac{1}{2} -b}~~ \bm{(}~\langle \rangle \langle ~\bm{)}  \\
\bm{(} ~~ \rangle~~\bm{)} ~~\xrightleftharpoons[\frac{1}{2} -b]{\frac{1}{2} +b}~~ \bm{(}~\rangle \langle \rangle ~\bm{)}  
\label{Eq_bend_rules_1}
\end{eqnarray}
Whereas, when a bend is enclosed by $) ($ interfaces, the transition rates are given by
\begin{eqnarray}
\nonumber
\bm{)} ~~ \langle~~\bm{(} ~~\xrightleftharpoons[\frac{1}{2} -b']{\frac{1}{2} +b'}~~ \bm{)}~\langle \rangle \langle ~\bm{(}  \\
\bm{)}~~ \rangle~~\bm{(} ~~\xrightleftharpoons[\frac{1}{2} +b']{\frac{1}{2} -b'}~~ \bm{)}~\rangle \langle \rangle ~\bm{(}
\label{Eq_bend_rules_2}
\end{eqnarray}
We note that the above rules apply even when the enclosing brackets are adjacent to the bracket being updated. In this case part of the resulting triad falls outside the enclosing brackets.

\subsection*{Irreducible Sequences}

\begin{figure}[t!]
\includegraphics[scale=0.24]{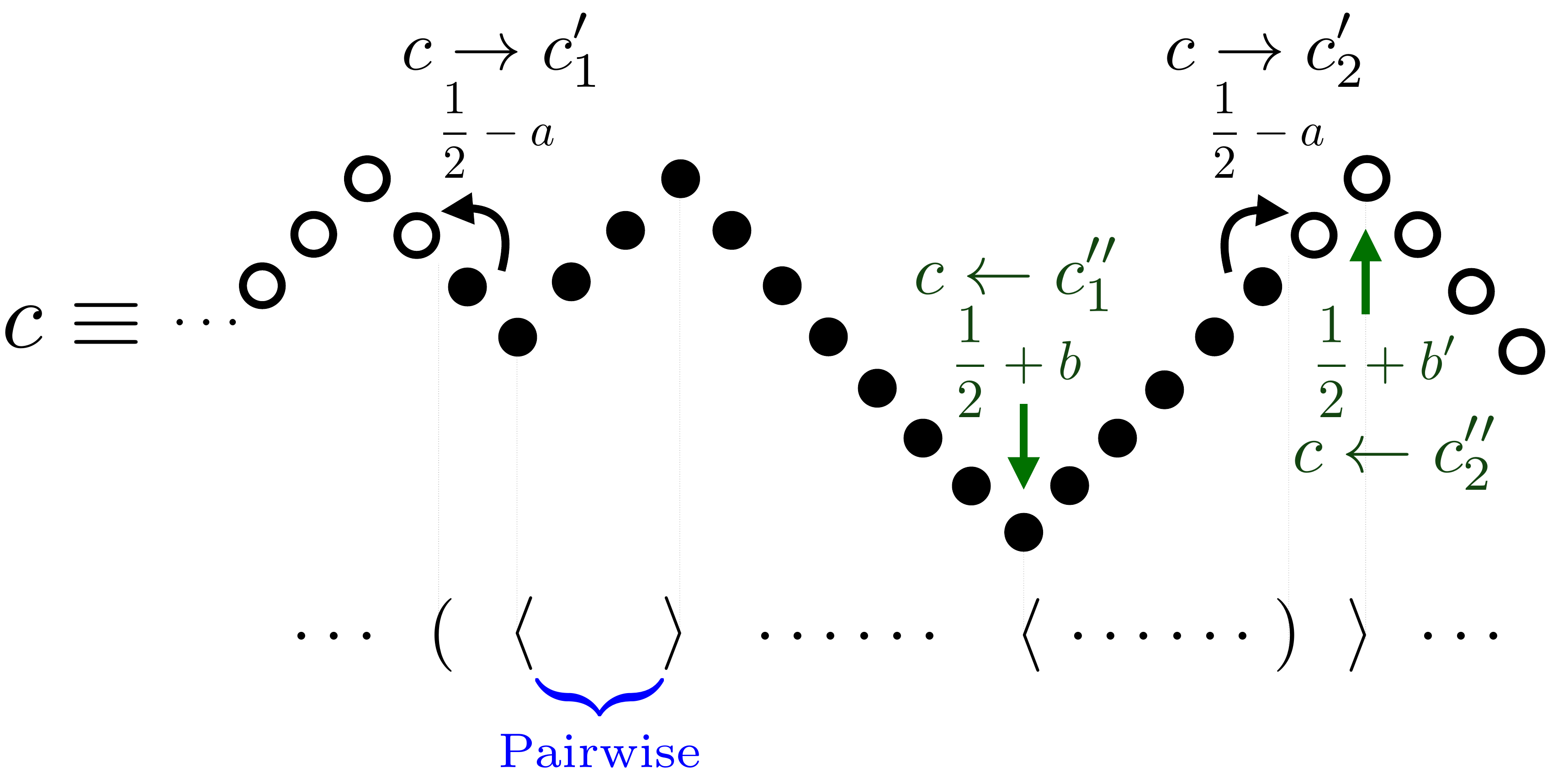}
\caption{A schematic illustration of the bunchwise balance condition in the LH model. The black arrows represent transitions out of the configuration $c$, while the green arrows represent transitions into the configuration. In this case, the transition rates to the configurations $c_1'$ and $c_2'$ are balanced by incoming transitions from $c_1''$ and $c_2''$ when the condition $2 a + b + b'$ is satisfied. The configuration in terms of interfaces and bends is represented below. Transitions between uninterrupted pairs of bends (or interfaces) are pairwise balanced. This sequence is an irreducible sequence of type $+ \equiv ( \langle ) \rangle$.
}
\label{bunchwise_balance_figure}
\end{figure}

Consider a general configuration of the LH model as shown in Fig.~\ref{bunchwise_balance_figure}, described by both density and tilt variables $\{\{ \sigma \},\{\tau\}\}$ and correspondingly by both interface and bend variables  $\{\{ \mathcal{I} \}, \{ \mathcal{B} \}\}$ as
\begin{equation}
c \equiv \hdots ~( ~~\langle ~~\rangle ~.~. ~. ~( ~~ )  ~.~. ~.~ \langle ~~) ~\rangle \hdots
\label{Eq_config_c4}
\end{equation}

In order to classify the transitions of this configuration into groups, we begin at the origin and  look at the enclosing interfaces and bends.
Suppose traversing rightwards, the first bracket encountered is $\langle$, we move forward until the next $\rangle$ is encountered. Whereas if the first bracket encountered is $\rangle$, we move backward until the next $\langle$ is encountered. Any bracket encountered during this traversal is added to the group. For example it is possible to encounter an interface: $($ or $)$. We apply this procedure recursively moving forward for a $($ bracket, and backward for $)$ brackets, adding any interface or bend brackets encountered into the group.
This procedure terminates when {\it all} the interface and bend brackets that are encountered have been paired, i.e. are closed by a corresponding bracket within the group. We note that these groups can contain sequences of any length such as $\langle ()() \hdots()() \rangle$, since once an open bracket $\langle$ is encountered, the group of transitions is not closed until the corresponding $\rangle$ is encountered. Similarly, the rest of the interfaces and bends in the configuration can be uniquely classified into groups in this manner.
Once again, this classification is unique, as every bracket belongs to the {\it smallest} group of closed brackets that contain it.

We next show that the transition rates into and out of each such group are balanced when $2 a + b + b' = 0$.  This occurs via two mechanisms: pairwise balance and bunchwise balance.
We begin by noticing that any uninterrupted pair of interfaces $(...)$ or bends $\langle ... \rangle$ obeys pairwise balance within itself. The proof for both these cases proceeds exactly as for the ASEP described above. Therefore, within each group, we can balance the transitions between these {\it reducible} pairs for any value of the transition rates $a, b$ and $b'$. Next, consider a general group of transitions represented by its sequence of interfaces and bends, for example $ \langle ...(...\rangle... ()()() ... \langle \rangle \langle \rangle... )$. We can reduce this sequence by pairing and eliminating the brackets representing transitions that satisfy pairwise balance, i.e. uninterrupted sequences of $()$ and $\langle \rangle$. Proceeding in this manner, we arrive at an {\it irreducible} sequence  $ \langle ...(...\rangle... )$. It is easy to see that there are only two types of irreducible sequences that arise, which we label $+$ and $-$ 
\begin{eqnarray}
\nonumber
+ \equiv ( \hdots \langle \hdots ) \hdots \rangle\\
- \equiv \langle \hdots ( \hdots \rangle  \hdots )
\label{plus_minus_definitions}
\end{eqnarray}
These represent groups of transitions that do not balance pairwise within themselves. We show below that these transitions are instead balanced as a group of four. Let us focus on the first case $ + \equiv ( \hdots \langle \hdots ) \hdots \rangle$, which is the case represented in Fig.~\ref{bunchwise_balance_figure}. The proof for the $-$ case proceeds analogously. 
Using Eq.~(\ref{Eq_interface_rules_2}), the transition out of the first bracket $c \rightarrow c_1'$ occurs with a rate $\frac{1}{2}-a$. Using Eq.~(\ref{Eq_bend_rules_1}), the reverse transition corresponding to the second bracket $c \leftarrow c_1''$ occurs with a rate $\frac{1}{2}+b$. Using Eq.~(\ref{Eq_interface_rules_1}), the transition out of the third bracket $c \rightarrow c_2'$ occurs with a rate $\frac{1}{2}-a$. Finally, using Eq.~(\ref{Eq_bend_rules_2}), the reverse transition corresponding to the last bracket $c \leftarrow c_2''$ occurs with a rate $\frac{1}{2}+b'$. The net incoming and outgoing transition rates between these pairs of forward and reverse transitions are given by
\begin{equation}
 r_{c \leftarrow c_1''}  +  r_{c \leftarrow c_2''} - r_{c\rightarrow c_1'} - r_{c\rightarrow c_2'} = 2 a + b + b'.
\end{equation}
Thus the net {\it incoming} transition rate from these four transitions is $R = 2 a + b + b'$. It is easy to show that the four corresponding reverse transitions also yield a net incoming transition rate of $R$. Since reducible pairs of brackets are pairwise balanced, these two groups of four represent all the unbalanced transitions associated with an irreducible sequence. Therefore the net incoming transition rate for a sequence of type $+$ is $2R$. Proceeding analogously for the type $-$ sequences, the net incoming rate can be shown to be $-2R$. We can interpret this asymmetry as follows, starting from a configuration with equal numbers of $+$ and $-$ irreducible sequences, for $R > 0$ the system  evolves towards a state in which the number of $+$ sequences is larger than the $-$ sequences in the steady state. Thus when $R >0$, the $+$ sequences are `favored', whereas the $-$ sequences are `unfavored'. This scenario is reversed when $R < 0$.
Importantly, when the condition $R = 2 a + b + b' = 0$ is satisfied, the transition rates for both types of irreducible sequences balance as
\begin{equation}
r_{c\rightarrow c_1'} + r_{c\rightarrow c_2'} =  r_{c \leftarrow c_1''}  +  r_{c \leftarrow c_2''}.
\end{equation}
Therefore when $R = 2 a + b + b' = 0$, the LH model satisfies the rate balance in Eq.  (\ref{Eq_steady_state_rate_balance}) and thus all configurations occur with equal probability in steady state.
Therefore this leads to the bunchwise balance condition for the probability currents in steady state
\begin{equation}
j_{c\rightarrow c_1'} + j_{c\rightarrow c_2'} =  j_{c \leftarrow c_1''}  +  j_{c \leftarrow c_2''},
\label{eq_bunchwise_balance_LH}
\end{equation}
which is a special case of the general bunchwise balance condition given in Eq.~(\ref{eq_bunchwise_balance}), with bunches containing two configurations each.
Finally, it is also easy to show that when $b = b' = -a$, the transitions within this group obey pairwise balance, as balance can now be achieved by pairing one of the interfaces and one of the bends that specify each irreducible string.

The equiprobable steady state when bunchwise balance is satisfied, leads to a product measure state in the thermodynamic limit. Product measure implies that local configurations occur with a probability independent of their neighbourhood in the steady state.
The bunchwise balance condition in Eq.~(\ref{bunchwise_balance_condition}) exhausts all possibilities in the LH parameter space for product measure in the steady state. 
This is confirmed by an exact calculation in Section \ref{exact_S_section}, where we have enumerated every possible lattice transition, starting from a totally disordered  (product measure) initial condition. We point out that a recent submission uses bunchwise balance (called multibalance in that paper) to find the steady states of several lattice models \cite{Indranil2020Multibalance}.


\section{Local Cross-correlation Function}
\label{S_section}

In this Section, we introduce the local cross-correlation function
$S$ and its disorder averaged version $\mathcal{S}$,
which measures the average correlation across the lattice between
particle (or tilt) sites with the gradients of their neighboring
tilt (or particle) sites. Although it is a local quantity, $S$ captures
important aspects of the early time dynamics as well as later time
dynamics, including coarsening towards ordered steady state phases. 

In terms of the discrete-space lattice variables for particles and
tilts, the local cross-correlation between the particle and tilt sub-lattices
can be expressed in two equivalent forms: $S_{\sigma\nabla\tau}\equiv S_{\tau\nabla\sigma}$. In any configuration $\{\sigma_{j},\tau_{j+{\scriptscriptstyle \frac{1}{2}}}\}$, we define the two forms at any time $t$ as
\begin{eqnarray}
\nonumber
S_{\sigma\nabla\tau}(t) & = & \sum_{j=1}^{N_{\text{sys}}} \frac{1}{2}\left(\tau_{j-\frac{1}{2}}-\tau_{j+\frac{1}{2}}\right)\sigma_{j},\\
S_{\tau\nabla\sigma}(t) & = & \sum_{j=1}^{N_{\text{sys}}} \frac{1}{2}\left(\sigma_{j+1}-\sigma_{j}\right)\tau_{j+\frac{1}{2}}.
\label{eq: S_defn}
\end{eqnarray}
The form $S_{\sigma\nabla\tau}$ measures the cross-correlation
between triads comprised by an individual particle and its two neighboring
tilts. Likewise, the other form $S_{\tau\nabla\sigma}$ measures the
cross-correlation between an individual tilt and its two neighboring
particles. The two forms are always equal for any lattice configuration
with periodic boundary conditions. This can be easily seen by re-ordering
the pairs of constituent $\sigma-\tau$ product terms typified in
Eq.~(\ref{eq: S_defn}). We represent either $S_{\sigma\nabla\tau}$ or $S_{\tau\nabla\sigma}$ at time $t$ by $S \equiv S(t)$. We next define the {\it disorder averaged} correlation function $\mathcal{S} \equiv \mathcal{S}(t)$ as
\begin{equation}
\mathcal{S}(t)\,=\, \frac{1}{N_{\text{sys}}}\left\langle S_{\sigma\nabla\tau}(t)\right\rangle \,=\, \frac{1}{N_{\text{sys}}} \left\langle S_{\tau\nabla\sigma}(t)\right\rangle,
\label{eq_S_definition}
\end{equation}
where the average $\langle \rangle$ is performed over multiple evolutions starting from different disordered initial configurations. 
Across the lattice at any instant in time, $S_{\sigma\nabla\tau}$
counts triads $\diagdown_{\bullet}\diagup$ and $\diagup^{\circ}\diagdown$
as $+1$, and counts triads $\diagdown_{\circ}\diagup$ and $\diagup^{\bullet}\diagdown$
as $-1$. Likewise, the triads $^{\circ}\diagdown_{\bullet}$ and
$_{\bullet}\diagup^{\circ}$ are counted by $S_{\tau\nabla\sigma}$
as $+1$, and triads $^{\bullet}\diagdown_{\circ}$ and $_{\circ}\diagup^{\bullet}$
are counted as $-1$. When the lattice update parameters $a,\,b$
and $b'$ are all positive, triads counted by the two forms as $+1$
are kinetically `favored', and occur more frequently; whereas triads
counted as $-1$ are kinetically `unfavored', occurring relatively
less frequently. Therefore for $a,\,b,\,b'>0$, $\mathcal{S}(t)$ evolves as
a positive valued function; its magnitude can be interpreted as a
measure of the relative local `satisfaction' of the favored triads
at time $t$. On the other hand, if either or both $b,\,b'$ are negative,
there is a possibility that $\mathcal{S}(t)$ can evolve as a negative valued
function of $t$, since the triads counted as $-1$ will now occur
in relatively greater numbers.

At this point, it is important to note an interesting connection between $S$, and the irreducible sequences defined in Section \ref{bunchwise_section}. We have
\begin{equation}
S = 2 N_{+} -  2 N_{-},
\label{eq_S_irreducible_connection}
\end{equation}
where $N_{+}$ and $N_{-}$ denote the number of irreducible sequences of types $+ \equiv (\langle)\rangle$ and $- \equiv \langle(\rangle)$ respectively. This can be shown easily as reducible pairs do not contribute to $S$.  

\begin{figure}[t]
\includegraphics[scale=0.26]{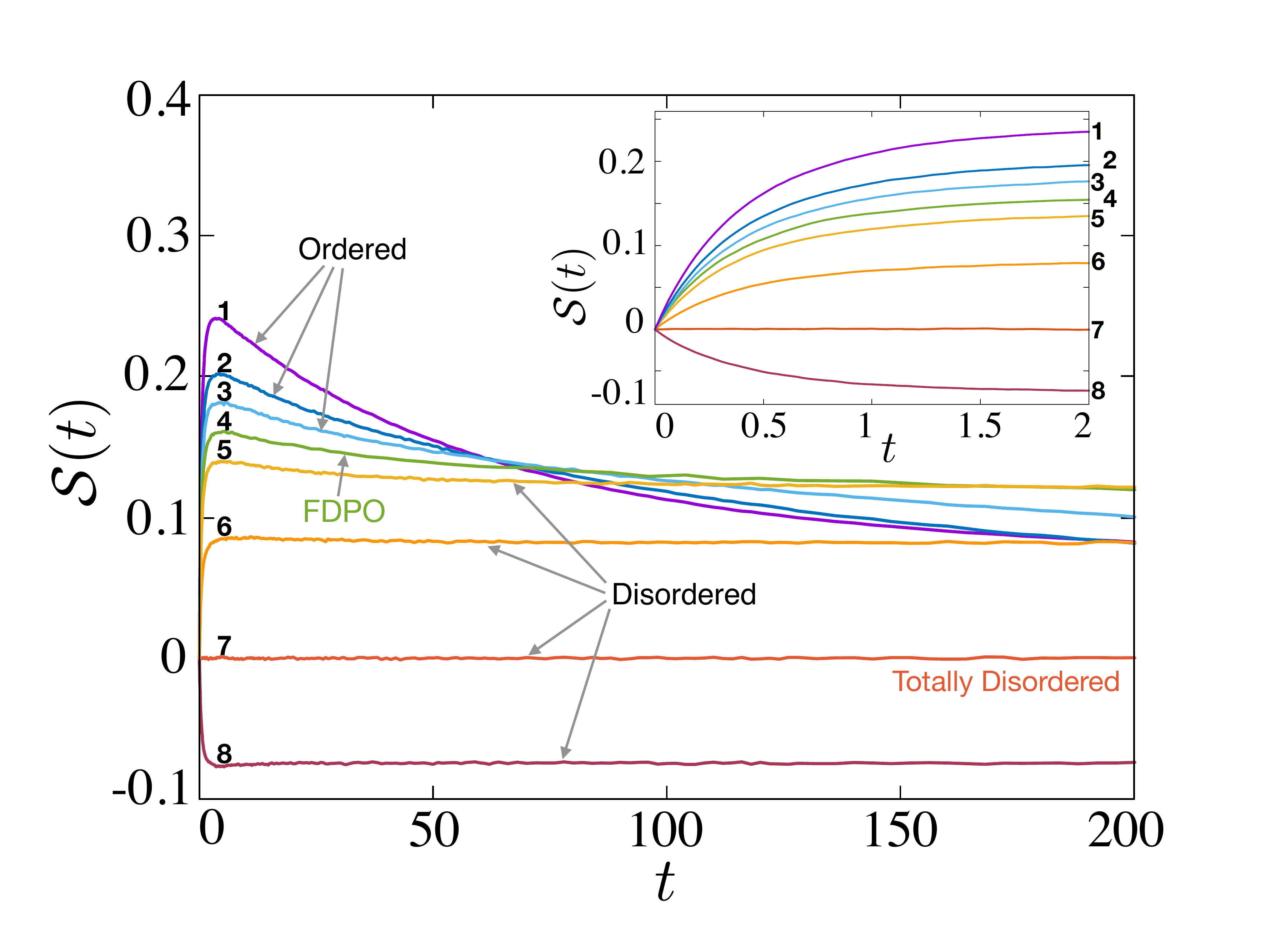}
\caption{Early time evolution of $\mathcal{S}$ leading to the ordered, disordered,
and FDPO steady state phases, beginning from an initial ensemble of random particle and
tilt configurations (totally disordered state). Plots {\bf 1}, {\bf 2} and {\bf 3} lead
to SPS, IPS and FPS phases in steady state, respectively. Plot {\bf 4} is
the order-disorder boundary (FDPO), and plots {\bf 5}, {\bf 6}, {\bf 7} and {\bf 8} correspond
to the disordered phase. In Plot {\bf 7}, $\mathcal{S}(t)$ does not evolve suggesting
that the completely disordered state persists up to and includes
the steady state. In this case, parameters $a,b,b'$  satisfy the relation $2a + b + b'=0$, derived using the condition of bunchwise balance in Section \ref{bunchwise_section}. {\bf Inset}: Expanded portion showing the growth of $\mathcal{S}(t)$ at small times. The system size used here is $N_{\mathrm{sys}}=512$. The parameter values used for ($a,b,b'$) are respectively: {\bf 1} ($0.4, 0.2, 0.2$), {\bf 2} ($0.4, 0.2, 0$), {\bf 3} ($0.4, 0.2, -0.1$), {\bf 4} ($0.4, 0.2, -0.2$), {\bf 5} ($0.4, 0.2, -0.3$), {\bf 6} ($0.3, -0.1, -0.1$), {\bf 7} ($0.3, -0.3, -0.3$) and {\bf 8} ($0.3, -0.5, -0.5$).
}
\label{fig:Early-time-evolution}
\end{figure}

\begin{figure}[t]
\includegraphics[scale=0.265]{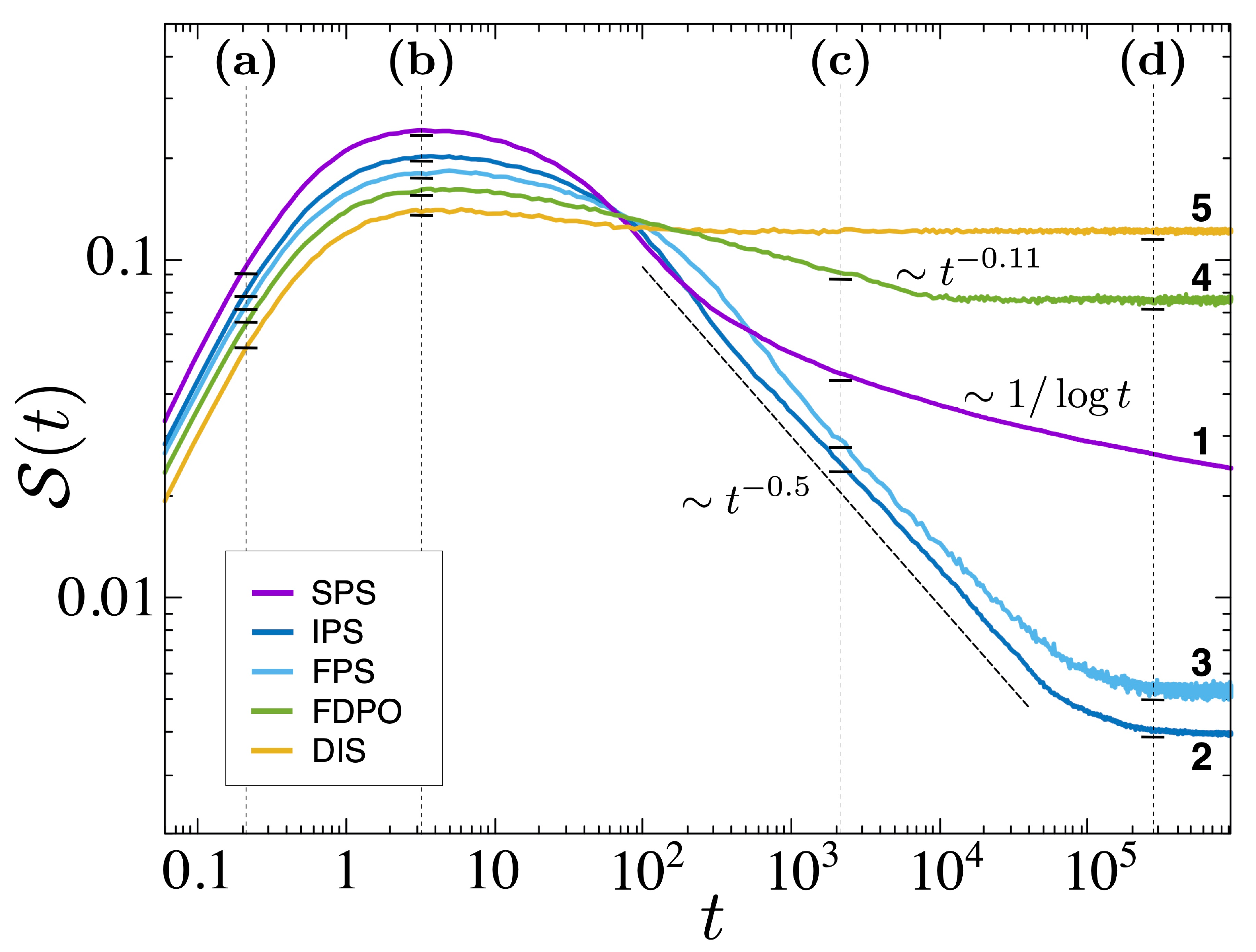}
\caption{Comparison of the early and late time behavior of \emph{$\mathcal{S}$} as the system evolves towards different steady state phases beginning from a random initial state. The parameter values chosen here are the same as in plots {\bf 1}-{\bf 5} in Fig.~\ref{fig:Early-time-evolution}.
The plots
{\bf 1}, {\bf 2}, {\bf 3} and {\bf 4} lead to SPS, IPS, FPS and FDPO phases respectively.
The different features shown by $\mathcal{S}(t)$ have been labeled as \textbf{(a)} early rise \textbf{(b)} broad maximum \textbf{(c)} decay during coarsening and \textbf{(d)} saturation at steady state. At late times, the system coarsens very slowly for SPS leading to $\mathcal{S}(t)\sim1/\log t$, whereas for FPS and IPS, the coarsening proceeds faster and
$\mathcal{S}(t)$ decays as a power law $\sim t^{-\phi}$ with $\phi\simeq0.5$.
For FDPO, $\mathcal{S}(t)$ decays with $\phi\simeq0.11$. In the cases of IPS,
FPS and FDPO, the steady state is reached, and $\mathcal{S}(t)$ reaches a saturation
value $\mathcal{S}_{ss}$, which decreases as the system size increases. In
the case of SPS, the decay of $\mathcal{S}(t)$ is so slow that the steady state
is not reached in the time of the simulation. In plot {\bf 5}, $\mathcal{S}(t)$ evolves to a disordered
steady state which saturates at a finite value at long times. The system size used here is $N_{\mathrm{sys}}=512$. 
\label{fig:Different-features-of}}
\end{figure}

The form of the function $\mathcal{S}(t)$ captures important aspects of the evolution from the initial state to the final steady state. In this paper, we choose the initial state to be totally disordered; the final steady state can be ordered, disordered, or the separatrix between ordered and disordered, depending on the values of $b$ and
$b'$, for $a>0$ as detailed in Section \ref{sec:The-Light-Heavy-Model}. For the totally disordered
initial state, we have $\mathcal{S}(0)=0$, since all forms of triads (favored
or unfavored) are equally likely. 
Below we provide an overview of the behavior of $\mathcal{S}(t)$ in the ordered
and disordered regimes, bringing out also the distinctive features at
early and late times.

\emph{Ordered Regime}: As discussed in Section \ref{sec:The-Light-Heavy-Model}, there are several
types of ordered phases. The light and heavy particles in all these ordered steady states are phase separated, and typical configurations
have large stretches with `inactive' triads such as $^{\bullet}\diagdown_{\bullet}$
or $_{\diagup}\circ\,^{\diagup}$, which are constituted by pairs
of identical particles or tilts, and do not contribute to $S(t)$.
Typically, there are only a small number $A \sim O(1)$ of active triads (that can be updated), and these occur at the cluster boundaries. Such triads are the only contributors to the dynamics of the system, and also to $S(t)$.
For all ordered phases
of a finite sized system, the saturation value is $\mathcal{S}_{ss}\approx A/N_{\text{sys}}$,
which vanishes in the thermodynamic limit $N_{\text{sys}}\rightarrow\infty$.
Below, we independently discuss the distinct steady state phases (SPS, IPS and FPS) within the ordered regime.

\emph{SPS}: The function $\mathcal{S}(t)$ rises linearly with a positive slope
at early time, till about $t\simeq0.5$ (plot 1 in Fig.~\ref{fig:Early-time-evolution}).
The linear growth results from the swift settling of particles and
tilts into locally satisfied triads, momentarily unhindered by exclusion
interactions. Over time, the effects of exclusion become more prominent,
and compete with the fulfillment of local satisfaction, at which point
the linear profile of $\mathcal{S}$ flattens to attain a broad maximum. At
later times $t > 100 $, exclusion effects promote the formation
of quasi-stable structures across the lattice, whose slow relaxation,
i.e. coarsening behavior towards a steady state, now dominates the
dynamics. As will be discussed further in Section \ref{section_late_time_S}, $S$ is a local
operator which is a global counter of these coarsening structures.
We see from plot 1 in Fig.~\ref{fig:Different-features-of} that
$\mathcal{S}$ decays as $\sim1/\log t$ at later times, reflecting the slow
coarsening process, which proceeds by activation in SPS. In the steady
state of a finite sized system, the number of active triads is $A=2$,
constituted by $\diagdown\bullet\diagup$ and $\diagup\circ\diagdown$.
Hence, we expect the SPS profile to saturate in the steady state at
$\mathcal{S}_{ss}\simeq2/N_{\text{sys}}$. Due to the slow logarithmic relaxation however,
we do not observe saturation over the time scales shown in Fig.~\ref{fig:Different-features-of}.

\emph{IPS}: At short times, $\mathcal{S}$ behaves as in SPS, growing linearly
and reaching a maximum (plot 2 in Fig.~\ref{fig:Early-time-evolution}).
Following the maximum, $\mathcal{S}(t)$ decays as a power-law $\sim t^{-\phi}$
with $\phi\simeq0.5$ (plot 2 in Fig.~\ref{fig:Different-features-of}).
For a finite sized lattice, we observe that $\mathcal{S}(t)$ saturates to a
constant, finite value $\mathcal{S}_{ss}\approx A/N_{\text{sys}}$, with $A\simeq2$.

\emph{FPS}: In this phase $\mathcal{S}$ behaves as in the IPS, growing linearly
at short times (plot 3 in Fig.~\ref{fig:Early-time-evolution}),
and decaying as a power law with $\phi\simeq0.5$ (plot 3 in Fig.~\ref{fig:Different-features-of}). In steady state, the saturation
value of $\mathcal{S}(t)$ is larger than in IPS: $\mathcal{S}_{ss}\approx A/N_{\text{sys}}$,
with $A\simeq2.5$.

\emph{FDPO}: This surface is the separatrix between the ordered and
disordered phases in the parameter space of the system. Here $\mathcal{S}$
grows linearly and attains a maximum (plot 4 in Fig.~\ref{fig:Early-time-evolution})
as in the ordered phases discussed above. Further $\mathcal{S}(t)$ shows a
slow decay with $\phi\simeq0.11$ (plot 4 in Fig.~\ref{fig:Different-features-of})
to a steady state with fluctuating long-range order \cite{Das2000fdpoPRL, Das2001fdpoPRE,chatterjee2006passive-sliders, kapri2016op-scaling}. For a finite system, the number of active
triads in steady state depends on the system size: $A\sim N_{\text{sys}}^{1-\mu}$.
The saturation value is thus $\mathcal{S}_{ss}\approx A/N_{\text{sys}}\sim N_{\text{sys}}^{-\mu}$ with $\mu \simeq 0.17$,
which tends to zero in the thermodynamic limit.

\emph{Disordered Regime}: Initially $\mathcal{S}(t)$ grows linearly with $t$, but the slope can be positive, negative or zero (plots 5-8 in Fig.~\ref{fig:Early-time-evolution}), depending on the sign of the combination $2a + b + b'$. Further, in this regime the extrema are less prominent since there is no coarsening. Particularly significant is the zero slope case in plot 7 of Fig.~\ref{fig:Early-time-evolution}, which corresponds to $d\mathcal{S}/dt = 0$, implying that $\mathcal{S}(t)$ does not evolve. This suggests that the steady state is completely disordered if $R = 2a + b + b'$ is zero, as was proved in Section \ref{bunchwise_section} using the condition of bunchwise balance. When $R$ is nonzero, $\mathcal{S}$ relaxes to a constant steady state value $\mathcal{S}_{ss}$ (plot 5 in Fig.~\ref{fig:Different-features-of}). Recall that in the disordered phase, the settling tendency of an \emph{H} particle in a local valley or an \emph{L} particle on a hill (at a rate set by $a$) is opposed by the unsettling of the valley or hill as soon as the particle arrives (set by $-b$ or $-b'$). Qualitatively, the value of $\mathcal{S}_{ss}$ is a measure of settling: $\mathcal{S}_{ss} > 0$ for $R > 0$ indicates that the majority of $H$ and $L$ particles find themselves well settled; likewise, for $R < 0$ we observe $\mathcal{S}_{ss} < 0$, implying that the majority are unsettled in this case. Close to $R=0$, we have checked that the value of $\mathcal{S}_{ss}$ is proportional to $R$. However, it varies in a strongly nonlinear manner and vanishes continuously as parameter values approach the transition (FDPO) locus. Further, the time taken to reach the constant value increases and appears to diverge as the locus of transitions is approached. This behavior is consistent with the hypothesis of a diverging correlation length at a mixed order transition, as discussed in Section \ref{sec:The-Light-Heavy-Model}.


\section{Exact Calculation of Early Time Slope of $\mathcal{S}(t)$}
\label{exact_S_section}


\begin{table*}[t]
\centering{}%
\begin{tabular}[t]{ccccccc}
\toprule 
 & $\quad$ & Transition (T) & $\quad$ & Transition Rate & $\quad$ & Contribution to $\Delta\mathcal{S}/\Delta t$\tabularnewline
\midrule
\midrule 
\addlinespace[0.1cm]
(1) &  & \multirow{1}{*}{$\:^{\bullet}\diagdown_{\circ}\:\longrightarrow\:^{\circ}\diagdown_{\bullet}$} &  & $\left(\frac{1}{2}+a\right)$ &  & $+\left(\frac{1}{2}+a\right)\frac{1-\sigma_{0}^{2}}{4}$\tabularnewline\addlinespace[0.1cm]
\midrule 
\addlinespace[0.1cm]
(2) &  & $\:^{\circ}\diagdown_{\bullet}\:\longrightarrow\:^{\bullet}\diagdown_{\circ}$ &  & $\left(\frac{1}{2}-a\right)$ &  & $-\left(\frac{1}{2}-a\right)\frac{1-\sigma_{0}^{2}}{4}$\tabularnewline\addlinespace[0.1cm]
\midrule 
\addlinespace[0.1cm]
(3) &  & $\:_{\circ}\diagup^{\bullet}\:\longrightarrow\:_{\bullet}\diagup^{\circ}$ &  & $\left(\frac{1}{2}+a\right)$ &  & $+\left(\frac{1}{2}+a\right)\frac{1-\sigma_{0}^{2}}{4}$\tabularnewline\addlinespace[0.1cm]
\midrule 
\addlinespace[0.1cm]
(4) &  & $\:_{\bullet}\diagup^{\circ}\:\longrightarrow\:_{\circ}\diagup^{\bullet}$ &  & $\left(\frac{1}{2}-a\right)$ &  & $-\left(\frac{1}{2}-a\right)\frac{1-\sigma_{0}^{2}}{4}$\tabularnewline\addlinespace[0.1cm]
\midrule 
\addlinespace[0.1cm]
(5) &  & $\:\diagup^{\circ}\diagdown\:\longrightarrow\:\diagdown_{\circ}\diagup$ &  & $\left(\frac{1}{2}-b'\right)$ &  & $-\left(\frac{1}{2}-b'\right)\frac{1-\sigma_{0}^{2}}{4}$\tabularnewline\addlinespace[0.1cm]
\midrule 
\addlinespace[0.1cm]
(6) &  & $\:\diagdown_{\circ}\diagup\:\longrightarrow\:\diagup^{\circ}\diagdown$ &  & $\left(\frac{1}{2}+b'\right)$ &  & \multicolumn{1}{c}{$+\left(\frac{1}{2}+b'\right)\frac{1-\sigma_{0}^{2}}{4}$}\tabularnewline\addlinespace[0.1cm]
\midrule 
\addlinespace[0.1cm]
(7) &  & $\:\diagdown_{\bullet}\diagup\:\longrightarrow\:\diagup^{\bullet}\diagdown$ &  & $\left(\frac{1}{2}-b\right)$ &  & $-\left(\frac{1}{2}-b\right)\frac{1-\sigma_{0}^{2}}{4}$\tabularnewline\addlinespace[0.1cm]
\midrule 
\addlinespace[0.1cm]
(8) &  & $\:\diagup^{\bullet}\diagdown\:\longrightarrow\:\diagdown_{\bullet}\diagup$ &  & $\left(\frac{1}{2}+b\right)$ &  & $+\left(\frac{1}{2}+b\right)\frac{1-\sigma_{0}^{2}}{4}$\tabularnewline\addlinespace[0.1cm]
\bottomrule
\addlinespace[0.2cm]
\end{tabular}

\caption{The various possible transitions associated with triads centered on a tilt site or a particle site, along with their respective
contributions to the early time slope of $d \mathcal{S}/dt$.
}
\label{table_transition}
\end{table*}

In this Section, we derive the exact early time behavior for the evolution of $\mathcal{S}$ within a short time interval $[0, \Delta t]$. We perform this computation starting from a totally disordered initial state. In this disordered state, the occupation probability of each site is independent of the others, with only an overall density and tilt imposed. For large systems this implies that the particle and tilt densities are statistically homogeneous across the lattice. This allows us to calculate the early time behavior exactly, by considering transitions at early time to be independent of each other. Finally we consider the contributions to $S$ from every possible transition out of a given configuration, and perform a disorder average over all initial conditions.


It is easy to show from the definition in Eq.~(\ref{eq_S_definition}) that starting with a totally disordered initial configuration at $t = 0$, we have $\mathcal{S}(t)=0$. For a given configuration $c$, the change in $S$ up to first order in time $\Delta t$ is given by
\begin{equation}
\Delta S = \sum_{\text{T}} p_{c}(\text{T}) \Delta S_{\text{T}},
\end{equation}
where $p_{c}(\text{T})$ represents the probability of occurrence of a particular transition $\text{T}$ in the configuration $c$, which we label by the state of the {\it triad} involved in the update. Note that these transitions $\text{T}$ only involve triads that can be updated. The sum is over all possible transitions out of the configuration. This form assumes that the initial updates are uncorrelated as they occur at a sufficient distance away from each other and therefore do not have any effect on each other. As the probability $p_{c}(\text{T}) \propto r_{\text{T}} \Delta t$, where $r_{\text{T}}$ is the microscopic rate of the transition, the probability of occurrence of two transitions within a short distance ($\le 3$) of each other leads to a higher order term of $O(\Delta t^2/N_{\text{sys}})$. 

To compute the change in the disorder averaged $\mathcal{S}$ to lowest order in time, we need to consider the neighbourhood of each triad that is updated. Since three sites are involved in every update, the sites immediately adjacent to the triad are also involved in the computation of $S$, i.e. we need to consider a  {\it quintet} associated with each update. We therefore consider all transitions associated with five consecutive sites in the system, and perform an average over all initial conditions. The change in $\mathcal{S}$ can then be written in the following form
\begin{equation}
 \Delta \mathcal{S} =\frac{1}{N_{\mathrm{sys}}}
\sum_{\text{T}} \langle p_c(\text{T})  \Delta S_{\text{T}} \rangle =
\frac{1}{N_{\mathrm{sys}}}\sum_{\text{T}} p(\text{T}) \langle \Delta S_{\text{T}} \rangle,   
\end{equation}
where $\langle \Delta S_{\text{T}} \rangle$ represents the {\it average} change in $S$ for a given transition $\text{T}$ over all possible initial conditions of the system. Above we have used the fact that for configurations drawn from the totally disordered state, the transition probabilities are independent of the configuration with $p_{c}(\text{T}) \equiv p(\text{T})$.
Next, we enumerate the contributions to $\Delta \mathcal{S}$ from all possible
transitions. The sum over these contributions
yields the exact early time behavior of $\mathcal{S}$.

\begin{figure}[t!]
\includegraphics[scale=0.26]{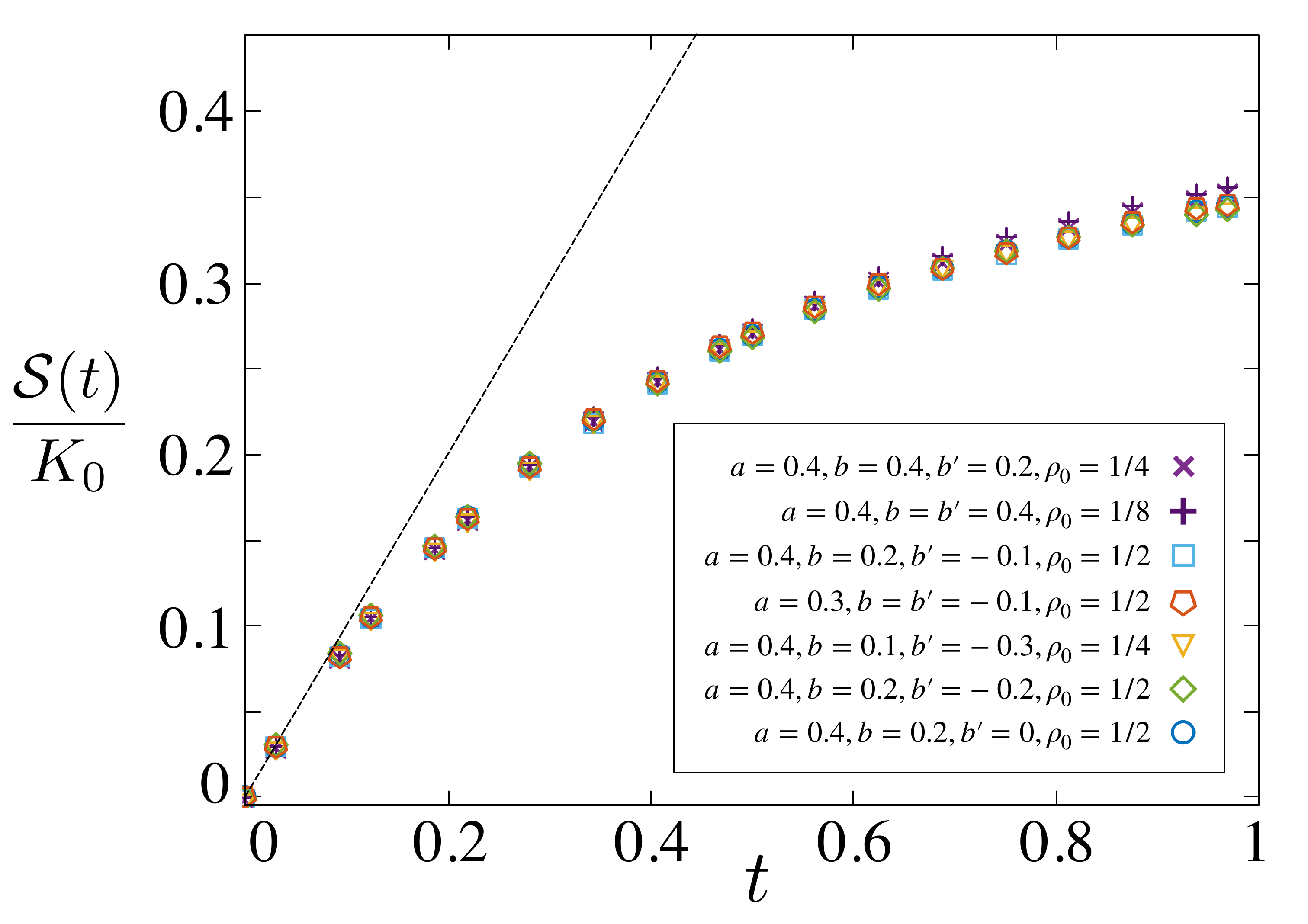}
\caption{Collapse of the early time evolution of $\mathcal{S}(t)/K_0$ for different values of parameters $a, b, b'$ and particle densities $\rho(\bullet)=\frac{1+\sigma_0}{2}$. Here $K_0=\frac{1-\sigma_0^{2}}{2}(2a+b+b')$ is the exact early time slope of $d\mathcal{S}/dt$ derived in Eq.~(\ref{eq_exact_early_time_slope}). The system size used here is $N_{\mathrm{sys}}=512$.
\label{S_Collapse}}
\end{figure}

As an example, we consider the contribution to $\mathcal{S}$ from a transition involving a particle exchange {\it across} a tilt, labeled as $\text{T} \equiv \boxed{{^\bullet}\diagdown_{\circ}}\:\longrightarrow\: \boxed{ ^{\circ}\diagdown_{\bullet}}$, which occurs at a rate $\left(\frac{1}{2} + a\right)$. We note that for any configuration $c$, the state of the triad and its position uniquely specify the transition, and we have dropped the position indices for brevity.
The probability of occurrence of such a triad in a disordered initial state can be computed from the individual occupation probabilities of each member of the triad. For a disordered configuration, we have $p({^\bullet}\diagdown_{\circ}) = p({\bullet})p(\diagdown)p({\circ})$. 
Next, the individual probabilities are given by their mean densities in the disordered (product measure) state
\begin{eqnarray}
\nonumber
p({\circ}) &=& \frac{1 - \sigma_0}{2},\\
\nonumber
p({\bullet}) &=& \frac{1 + \sigma_0}{2},\\
p(\diagdown) &=& \frac{1}{2}.
\end{eqnarray}
Since, this transition can occur anywhere in the system, the probability that this transition occurs in the first time step $[0,\Delta t]$ is 
\begin{equation}
p(\text{T}) = N_{\text{sys}} \left(\frac{1 - \sigma_0^2}{8}\right) \left(\frac{1}{2} + a\right) \Delta t.    
\end{equation}
Next, we consider the possible changes in $S$ that such a transition can cause. To do this, we need to consider the state of the system on the {\it quintet} associated with the transition site. It is easy to see that there are only three possibilities that produce a non-zero change to $S$ for such a transition, namely (i) $\diagup \boxed{{^\bullet}\diagdown_{\circ}} \diagup$ (ii) $\diagup \boxed{{^\bullet}\diagdown_{\circ}} \diagdown$ and (iii) $\diagdown \boxed{{^\bullet}\diagdown_{\circ}} \diagup$. The contributions from each of these cases to $S$ are
(i) $+4$ (ii) $+2$ and (iii) $+2$ respectively. Summing over all three cases we arrive at the contribution to $\mathcal{S} = \langle S \rangle/N_{\text{sys}}$ from this transition $\text{T} \equiv \boxed{{^\bullet}\diagdown_{\circ}}$, labeled as $(1)$
\begin{equation}
 \Delta \mathcal{S}_{(1)} = \left(\frac{1}{2}+a\right)\frac{\left(1-\sigma_{0}^{2}\right)}{4} \Delta t,
\end{equation}
as stated in the first row of Table \ref{table_transition}.
Next, all the transitions that can occur in the system can be treated in the same manner. There are eight possible states of a given triad, centered either on a particle site or a tilt site. The contributions from each of these cases have been summarized in Table \ref{table_transition}. Finally, summing over the contributions from all possible transitions, we arrive at the following exact behavior of $\mathcal{S}$ at early time
\begin{eqnarray}
 \Delta \mathcal{S}  =\frac{\left(1-\sigma_{0}^{2}\right)}{2}\left(2 a+ b+b' \right) \Delta t.
\label{eq_exact_early_time_slope}
\end{eqnarray}
We note that the slope of $d \mathcal{S}/ dt$ has the same factor $R = (2 a + b + b')$ that appears in the bunchwise balance condition, and therefore when the system exhibits bunchwise balance, the correlation $\mathcal{S}$ does not evolve in time. In Fig.~\ref{S_Collapse} we display the early time behavior of $\mathcal{S}$ obtained from simulations for various densities and microscopic rates in the LH model, displaying a linear rise and collapse consistent with Eq.~(\ref{eq_exact_early_time_slope}).

Finally we present an argument to show that $R =2a + b+b'=0$ is a necessary and sufficient condition for product measure in the steady state. We have shown in Section \ref{bunchwise_section} that $R=0$ implies product measure through bunchwise balance. To prove that product measure implies $R=0$, we assume the contrary, i.e. that $R$ is nonzero. However, we have derived the exact evolution of $\mathcal{S}$ starting from a product measure initial condition in Eq.~(\ref{eq_exact_early_time_slope}). We therefore have $\frac{d\mathcal{S}}{dt} \propto R$, which is nonzero.
Thus $\mathcal{S}$ must change from its initial value of $0$. But this is not possible if the state is product measure, as product measure states have $\mathcal{S}=0$. Therefore the condition $R=0$ implies product measure and the locus of the bunchwise balance condition exhausts all possibilities in the LH parameter space for product measure in the steady state.

\section{Early time evolution of $\mathcal{S}$ from mean field theory}
\label{linearized_mean_field_section}

In this Section, we study the time evolution of $\mathcal{S}(t)$ within a linearised
mean field approximation which neglects correlations between different sites. We will see that it describes the early time behavior very well, while at late times, it displays instabilities that identify the passage to ordered states.

Let us define fluctuation variables $\delta\sigma_{j}=\sigma_{j}-\sigma_{0}$
and $\delta\tau_{j+\frac{1}{2}}=\tau_{j+\frac{1}{2}}-\tau_{0}$ for the particles
and tilts respectively about their mean densities $\sigma_{0}$, $\tau_{0}$,
further restricting ourselves to the case of an equal number of up
and down tilts ($\tau_{0}=0$). In terms of $\delta\sigma$ and $\delta\tau$,
the function $\mathcal{S}(t)$ defined in Eq.~(\ref{eq_S_definition}) can be expressed
as
\begin{eqnarray}
\hspace{-0.5cm}
\mathcal{S}(t) & = & \frac{1}{N_{\text{sys}}}
\sum_{j=1}^{N_{\text{sys}}}
\frac{1}{2}\left\langle \left(\delta\sigma_{j+1}(t)-\delta\sigma_{j}(t)\right)\delta\tau_{j+\frac{1}{2}}(t)\right\rangle.
\label{eq: S_alt_defn}
\end{eqnarray}

Within the mean field approximation, the fluctuation variables
on different lattice sites are taken to be uncorrelated. Thus $\mathcal{S}$
can be approximated as $\frac{1}{N_{\text{sys}}}\sum_{j=1}^{N_{\text{sys}}}\,\frac{1}{2}\left\langle \delta\sigma_{{\scriptscriptstyle j+1}}-\delta\sigma_{{\scriptscriptstyle j}}\right\rangle \left\langle \delta\tau_{{\scriptscriptstyle j+1/2}}\right\rangle $.
Furthermore, with the expectation that non-linear effects would not
feature prominently at very early time, we focus on deriving the evolution
of $\mathcal{S}$ from the linearised mean field equations governing $\delta\sigma$
and $\delta\tau$.

In preceding studies \cite{LBR2000SPS,DasBasu2001Dynamic-Scaling,Chakraborty2017LH-statics,Chakraborty2019CoupledModes-LH},
the LH model has been analysed in the continuum, using hydrodynamic
mean field equations describing coarse grained density and tilt fields.
However, while the continuum approximation may be justified for large-distance,
long-time properties, it fails to describe the early time behavior
even qualitatively. Hence we deal with the linearised mean field equations
on a discrete lattice, and find that our results reproduce the principal
features of $\mathcal{S}(t)$, observed in simulations at early times, aside from
matching the initial slope which was exactly determined in Section \ref{exact_S_section}.

\subsection{Lattice mean field equations}

Recalling that the number of heavy particles and the number of up
tilts both are conserved, we write the following discrete continuity
equations 
\begin{eqnarray}
\partial_{t}\sigma_{j} & = & J_{\sigma}\left(j-1,j\right)-J_{\sigma}\left(j,j+1\right),\nonumber \\
\nonumber \\
\partial_{t}\tau_{j+\frac{1}{2}} & = & J_{\tau}\left(j-{\scriptstyle \frac{1}{2}},j+{\scriptstyle \frac{1}{2}}\right)-J_{\tau}\left(j+{\scriptstyle \frac{1}{2}},j+{\scriptstyle \frac{3}{2}}\right).
\label{eq: Continuity_eqns}
\end{eqnarray}
\noindent Here, the terms $J_{\sigma,\tau}\left(i,i+1\right)$ represent
the resultant particle or tilt currents from $i$ to $i+1$, where
$i$ stands for $j$ or $j+{\frac{1}{2}}$. The currents
$J_{\sigma,\tau}$ can be calculated within the mean field approximation. 
We have
\begin{widetext}
\begin{small}
\begin{eqnarray}
\nonumber
J_{\sigma}\left(j,j+1\right) & = & \frac{a}{2}\,\tau_{j+\frac{1}{2}}\left\{ \left(1+\sigma_{j}\right)\left(1-\sigma_{j+1}\right)+\left(1+\sigma_{j+1}\right)\left(1-\sigma_{j}\right)\right\} +\frac{\nu_{\sigma}}{2}\left(\sigma_{j}-\sigma_{j+1}\right)+\eta_{j}^{\sigma}(t), \\
\nonumber
J_{\tau}\left(j+{\scriptstyle \frac{1}{2}},j-{\scriptstyle \frac{1}{2}}\right) & = & \left(\frac{1+\sigma_{j}}{2}\left(b+b'\right)-b'\right)\left\{ \left(1+\tau_{j+\frac{1}{2}}\right)\left(1-\tau_{j-\frac{1}{2}}\right)+\left(1+\tau_{j-\frac{1}{2}}\right)\left(1-\tau_{j+\frac{1}{2}}\right)\right\} +\,\frac{\nu_{\tau}}{2}\left(\tau_{j+\frac{1}{2}}-\tau_{j-\frac{1}{2}}\right)+\eta_{j+\frac{1}{2}}^{\tau}(t),\\
\label{eq: J_sigma_tau}
\end{eqnarray}
\end{small}
\end{widetext}
where the terms $\frac{a}{2}\,\tau_{{\scriptscriptstyle j+\frac{1}{2}}}\,\{ \hdots \}$ and
$\left({\scriptscriptstyle \frac{1+\sigma_{j}}{2}}\left(b+b'\right)-b'\right)\{ \hdots \}$
are the `systematic' contributions to the currents originating from
the update rules. 
These rules also generate diffusive terms $\frac{\nu_{\sigma}}{2}\left(\sigma_{j}-\sigma_{j+1}\right)$
and $\frac{\nu_{\tau}}{2}\left(\tau_{{\scriptscriptstyle j+\frac{1}{2}}}-\tau_{{\scriptscriptstyle j-\frac{1}{2}}}\right)$;
their coefficients $\nu_{\sigma}$ and $\nu_{\tau}$ are proportional
to the frequency of particle and tilt updates at every time step.
Finally, the currents $J_{\sigma,\tau}$ also include phenomenologically
added noise terms $\eta^{\sigma}$ and $\eta^{\tau}$. We consider
the noise terms to be delta-correlated with the correlators $\left\langle \eta_{{\scriptscriptstyle j}}^{\sigma}(t)\:\eta_{{\scriptscriptstyle j}'}^{\sigma}(t')\right\rangle =D^{\sigma}\:\delta_{{\scriptscriptstyle j,j'}}\delta(t-t')$
and $\left\langle \eta_{{\scriptscriptstyle j+\frac{1}{2}}}^{\tau}(t)\:\eta_{{\scriptscriptstyle j'+\frac{1}{2}}}^{\tau}(t')\right\rangle =D^{\tau}\:\delta_{{\scriptscriptstyle j,j'}}\delta(t-t')$,
where $D^{\sigma}$ and $D^{\tau}$ are the respective strengths.

We insert the current expressions $J_{\sigma,\tau}$ in Eq.~(\ref{eq: J_sigma_tau})
into the continuity equations for $\sigma$ and $\tau$ [Eq.~(\ref{eq: Continuity_eqns})],
by retaining only the linear terms from the systematic parts of $J_{\sigma,\tau}$.
Re-expressing the resulting equations in terms of fluctuation variables
$\delta\sigma$ and $\delta\tau$, we arrive at the following linearised,
mean field equations
\begin{eqnarray}
\partial_{t}\delta\sigma_{j} & = & a(1-\sigma_{0}^{2})\left\{ \delta\tau_{j-\frac{1}{2}}-\delta\tau_{j+\frac{1}{2}}\right\} \nonumber \\
 &  & +\frac{\nu_{\sigma}}{2}\left\{ 2\delta\sigma_{j}-\delta\sigma_{j+1}-\delta\sigma_{j-1}\right\} +f_{j}^{\sigma}(t)
\label{eq: Evolution_eqn1}
\end{eqnarray}
and
\begin{eqnarray}
\partial_{t}\delta\tau_{j+\frac{1}{2}} & = & -\left(\frac{b+b'}{2}\right)\left\{ \delta\sigma_{j}-\delta\sigma_{j+1}\right\} \nonumber \\
 &  & +\frac{\nu_{\tau}}{2}\left\{ 2\delta\tau_{j+\frac{1}{2}}-\delta\tau_{j+\frac{3}{2}}-\delta\tau_{j-\frac{1}{2}}\right\} +f_{j+\frac{1}{2}}^{\tau}(t),\nonumber \\
\label{eq: Evolution_eqn2}
\end{eqnarray}
where $f^{\sigma}$ and $f^{\tau}$ are negative discrete gradients
of the noise terms: $f_{{\scriptscriptstyle {\scriptscriptstyle j}}}^{\sigma}=\eta_{{\scriptscriptstyle j-1}}^{\rho}-\eta_{{\scriptscriptstyle j}}^{\rho}$
and $f_{{\scriptscriptstyle j+\frac{1}{2}}}^{\tau}=\eta_{{\scriptscriptstyle j-\frac{1}{2}}}^{\tau}-\eta_{{\scriptscriptstyle j+\frac{1}{2}}}^{\tau}$.
It also proves expedient to define height fields for particles $\{h_{j}^{\sigma}\}$
 and tilts $\{h_{j+\frac{1}{2}}^{\tau}\}$ as follows
\begin{equation}
h_{j}^{\sigma}=\sum_{j'=1}^{j}\sigma_{j'}\quad \quad h_{j+\frac{1}{2}}^{\tau}=\sum_{j'=1}^{j}\tau_{j'+\frac{1}{2}}.
\end{equation}
Evidently, we have $\sigma_{{\scriptscriptstyle j}}=h_{{\scriptscriptstyle j}}^{\sigma}-h_{{\scriptscriptstyle j-1}}^{\sigma}$
and $\tau_{{\scriptscriptstyle j+\frac{1}{2}}}=h_{{\scriptscriptstyle j+\frac{1}{2}}}^{\tau}-h_{{\scriptscriptstyle j-\frac{1}{2}}}^{\tau}$.
The corresponding relations in discrete Fourier space are
\begin{eqnarray}
\widehat{\delta\sigma}_{k}(t) & = & \widehat{h}_{k}^{\sigma}(t)\left(1-e^{{\scriptscriptstyle ik}}\right),
\nonumber \\
\widehat{\delta\tau}_{k}(t) & = & \widehat{h}_{k}^{\tau}(t)\left(1-e^{{\scriptscriptstyle ik}}\right),
\label{eq: height_variables}
\end{eqnarray}
where we have defined the Fourier variables as $\widehat{\delta\sigma}_{k}=\sum_{j=1}^{N_{\text{sys}}}e^{ikj}\,\delta\sigma_{{\scriptscriptstyle j}}$
and $\widehat{\delta\tau}_{k}=\sum_{j=1}^{N_{\text{sys}}}e^{ik\left(j+{\scriptscriptstyle \frac{1}{2}}\right)}\,\delta\tau_{{\scriptscriptstyle j+{\scriptscriptstyle \frac{1}{2}}}}$;
$\widehat{h}_{k}^{\sigma}=\sum_{j=1}^{N_{\text{sys}}} e^{ikj}\,h_{\scriptscriptstyle{j}}^{\sigma}$
and $\widehat{h}_{k}^{\tau}=\sum_{j=1}^{N_{\text{sys}}}e^{ik\left(j+{\scriptscriptstyle \frac{1}{2}}\right)}\,h_{{\scriptscriptstyle j+{\scriptscriptstyle \frac{1}{2}}}}^{\tau}$
where $k=\frac{2\pi}{N_{\text{sys}}}\,m\:(m\in\mathbb{Z})$.

\subsection{Solving the linearized mean field equations }

The coupled, linearised equations [Eqs.~(\ref{eq: Evolution_eqn1})
and (\ref{eq: Evolution_eqn2})] can be solved by going to Fourier
space. We write them as a matrix equation in the following form
\begin{eqnarray}
\partial_{t}\left(\begin{array}{c}
\widehat{\delta\sigma}_{k}\\
\widehat{\delta\tau}_{k}
\end{array}\right) & = & \mathcal{M}\left(\begin{array}{c}
\widehat{\delta\sigma}_{k}\\
\widehat{\delta\tau}_{k}
\end{array}\right)+\left(\begin{array}{c}
\widehat{f}_{k}^{\sigma}\\
\widehat{f}_{k}^{\tau}
\end{array}\right),
\end{eqnarray}
where the diagonal and off-diagonal elements of matrix $\mathcal{M}$
involve the diffusive and drift terms respectively
\begin{eqnarray}
\mathcal{M} & = & \left[\begin{array}{cc}
2\nu_{\sigma}\sin^{2}\frac{k}{2} & i2a\left(1-\sigma_{0}^{2}\right)\sin\frac{k}{2}\\
-i\left(b+b'\right)\sin\frac{k}{2} & 2\nu_{\tau}\sin^{2}\frac{k}{2}
\end{array}\right].
\end{eqnarray}

It is straightforward to find the eigenmodes $\widehat{P}_{k}$ and
$\widehat{Q}_{k}$ of $\mathcal{M}$, and their corresponding eigenvalues
$\lambda_{k}^{\pm}$. If the update frequencies are equal for particles
and tilts, as in our simulations, then $\nu_{\sigma}=\nu_{\tau}=\nu$
is the effective diffusion constant for both the eigenmodes. The eigenvalue
expressions then reduce to 
\begin{eqnarray}
\lambda_{k}^{\pm} & = & 2\nu\sin^{2}\frac{k}{2}\pm c\sin\frac{k}{2},
\label{eq: eigenvalue}
\end{eqnarray}
and the corresponding eigenmodes are
\begin{eqnarray}
\nonumber
\widehat{P}_{k} & = & \frac{1}{\mathcal{C}}\widehat{\delta\sigma}_{k}+\widehat{\delta\tau}_{k},\\
\widehat{Q}_{k} & = & -\frac{1}{\mathcal{C}}\widehat{\delta\sigma}_{k}+\widehat{\delta\tau}_{k}.
\label{eq: eigenmodes}
\end{eqnarray}
The constants $c$ and $\mathcal{C}$ are related to the off-diagonal elements
of matrix $\mathcal{M}$, and are defined as $c=\sqrt{2a(1-\sigma_{0}^{2})(b+b')}$
and $\mathcal{C}=\sqrt{2a\left(1-\sigma_{0}^{2}\right)/(b+b')}$. Here, $\mathcal{C}$
is a proportionality constant which governs the relative admixture
of $\widehat{\delta\sigma}_{k}$ and $\widehat{\delta\tau}_{k}$ in
the eigenmodes. 

The constant $c$ changes in an important way across the phase boundary,
associated with the fact that it is real, imaginary or zero, according
to whether $b+b'$ is positive, negative or zero. Depending on whether
$c$ is real or imaginary, the system evolves into an ordered state
or a disordered state. In the ordered regime, $c$ is real and represents
an instability; fluctuations in particle and tilt densities grow into
instabilities which are curbed by non-linearities that are neglected
in the linearised theory. By contrast, in the disordered regime, $c$
is imaginary and its magnitude represents a speed; the fluctuations
do not grow, but move as mixed-mode kinematic waves with speeds given by the
magnitudes of the eigenvalues, i.e. $c$. The case when $c$ is zero
($b+b'=0$) defines the order-disorder phase boundary. In this case,
the linear coupling vanishes in Eq.~(\ref{eq: Evolution_eqn2}). Hence,
the tilt field evolves autonomously, directing the evolution of the
particle field --- an example of a passive scalar problem \cite{Kraichnan1994Passive-scalar,Falkovich2001Passive-scalar}.

The evolution equations of the eigenmodes $\widehat{P}_{k}$ and $\widehat{Q}_{k}$
are
\begin{eqnarray}
\partial_{t}\widehat{P}_{k} & = & \lambda_{k}^{+}\widehat{P}_{k}+\widehat{f}_{k}^{P},\nonumber \\
\partial_{t}\widehat{Q}_{k} & = & \lambda_{k}^{-}\widehat{Q}_{k}+\widehat{f}_{k}^{Q}.\\
\nonumber 
\end{eqnarray}
In Fourier space, we define the height fields $\widehat{h}_{k}^{P}$
and $\widehat{h}_{k}^{Q}$ for the eigenmodes, in analogy with $\widehat{h}_{k}^{\sigma,\tau}$
defined earlier just below Eq.~(\ref{eq: height_variables}) for particles
and tilts. Integrating the equations for $\widehat{h}_{k}^{P}$ and
$\widehat{h}_{k}^{Q}$, we find
\begin{eqnarray}
\widehat{h}_{k}^{P}(t) & = & \widehat{h}_{k}^{P}(0)e^{-\lambda_{k}^{+}t}+e^{-\lambda_{k}^{+}t}\int_{0}^{t}dt'\:e^{\lambda_{k}^{+}t'}\widehat{\eta}_{k}^{P}(t),\nonumber \\
\widehat{h}_{k}^{Q}(t) & = & \widehat{h}_{k}^{Q}(0)e^{-\lambda_{k}^{-}t}+e^{-\lambda_{k}^{-}t}\int_{0}^{t}dt'\:e^{\lambda_{k}^{-}t'}\widehat{\eta}_{k}^{Q}(t).
\label{eq: heqs}
\end{eqnarray}
The right hand side involves the initial condition and an
integration over the noise. For a randomly chosen initial configuration,
the initial particle and tilt profiles are delta-correlated, i.e.
$\left\langle \delta\sigma_{{\scriptscriptstyle j}}(0)\:\delta\sigma_{{\scriptscriptstyle j}'}(0)\right\rangle =M^{\sigma}\:\delta_{{\scriptscriptstyle j,j'}}$
and $\left\langle \delta\tau_{{\scriptscriptstyle j+\frac{1}{2}}}(0)\:\delta\tau_{{\scriptscriptstyle j'+\frac{1}{2}}}(0)\right\rangle =M^{\tau}\:\delta_{{\scriptscriptstyle j,j'}}$,
where $M^{\sigma}=1-\sigma_{0}^{2}$ and $M^{\tau}=1$ are the correlation
strengths computed for the random initial configuration. This leads
to $\left\langle \widehat{P}_{{\scriptscriptstyle k}}(0)\:\widehat{Q}_{{\scriptscriptstyle k'}}(0)\right\rangle =N_{\text{sys}}\:M^{PQ}\:\delta_{{\scriptscriptstyle k},{\scriptscriptstyle -k'}}$
with $M^{PQ}=\left(\frac{1}{\mathcal{C}^{2}}M^{\sigma}+M^{\tau}\right)$. The
correlator (i) $\left\langle \widehat{h}_{{\scriptscriptstyle k}}^{P}(0)\widehat{h}_{{\scriptscriptstyle k'}}^{Q}(0)\right\rangle $
can be derived from the eigenmode correlator $\left\langle \widehat{P}_{{\scriptstyle {\scriptscriptstyle k}}}(0)\:\widehat{Q}_{{\scriptstyle {\scriptscriptstyle k'}}}(0)\right\rangle $,
and the correlator (ii) $\left\langle \widehat{\eta}_{{\scriptscriptstyle k}}^{P}(t')\widehat{\eta}_{{\scriptscriptstyle k'}}^{Q}(t'')\right\rangle $
comprises a linear combination of the noise correlators for the fluctuation
variables; $\left\langle \eta_{{\scriptscriptstyle j}}^{\sigma}(t)\:\eta_{{\scriptscriptstyle j}'}^{\sigma}(t')\right\rangle $
and $\left\langle \eta_{{\scriptscriptstyle {j+\frac{1}{2}}}}^{\tau}(t)\:\eta_{{\scriptscriptstyle {j'+\frac{1}{2}}}}^{\tau}(t')\right\rangle $.
The expressions of correlators (i) and (ii) are respectively
\begin{eqnarray}
\left\langle \widehat{h}_{k}^{P}(0)\:\widehat{h}_{k'}^{Q}(0)\right\rangle  & = & \frac{N_{\text{sys}}\:M^{PQ}}{\left(1-e^{ik}\right)\left(1-e^{-ik'}\right)}\:\delta_{k,-k'},
\label{eq: heightcorrelator}\\
\nonumber \\
\left\langle \widehat{\eta}_{k}^{P}(t')\:\widehat{\eta}_{k^{'}}^{Q}(t'')\right\rangle  & = & N_{\text{sys}}\:D^{PQ}\:\delta_{k,-k'}\delta(t'-t''),
\label{eq: noisecorrelator}
\end{eqnarray}
\noindent where $D^{PQ}=\left(\frac{1}{\mathcal{C}^{2}}D^{\sigma}+D^{\tau}\right)$. 

\subsection{Evolution of $\mathcal{S}$}

\begin{figure*}[ht!]
\includegraphics[scale=0.45]{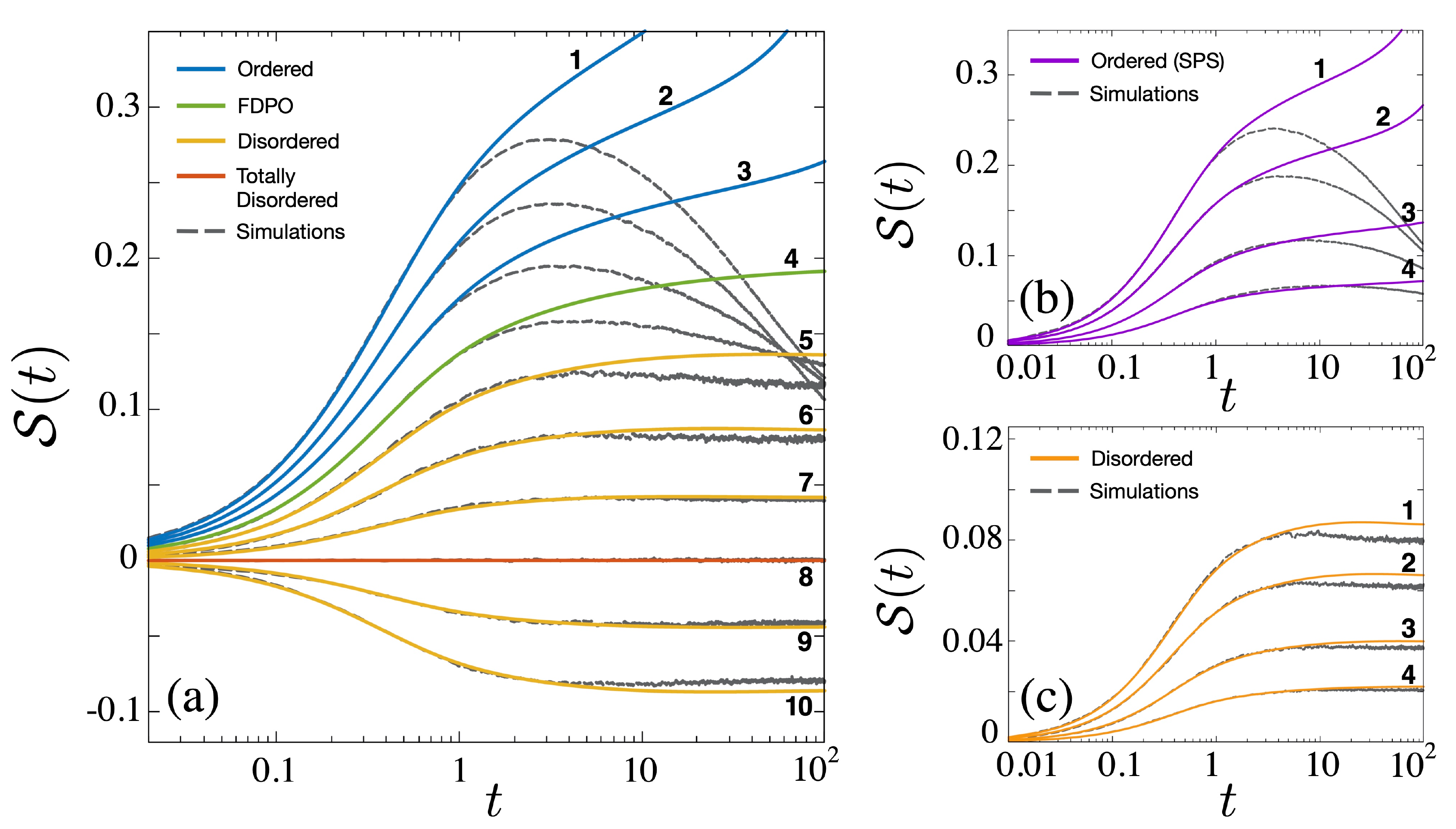}
\caption{Comparison of the early time evolution of $\mathcal{S}(t)$ derived from mean field theory (colored plots) with simulations (grey, dashed plots). Linear-log plots are shown for \textbf{(a)} varying parameters $a, b, b'$  across the phase diagram with mean particle density
$\rho(\bullet)=\frac{1+\sigma_0}{2} = 1/2$, for different ordered and disordered steady state phases, \textbf{(b)} varying particle densities $\rho(\bullet) = 1/2$, $1/4$, $1/8$  and $1/16$ in plots $ \mathbf{1}-\mathbf{4}$, with parameters $a=0.4$, and $b=b'=0.2$  fixed in the ordered regime (SPS), and \textbf{(c)} varying particle densities $\rho(\bullet) = 1/2$, $ 1/4$, $ 1/8$  and $1/16$ in plots $ \mathbf{1}-\mathbf{4}$, with parameters $a = 0.4, b = 0.2$ and $b' = -0.3$  fixed in the disordered regime. In all three cases \textbf{(a)}, \textbf{(b)} and \textbf{(c)},  the analytical evolution in Eq.~(\ref{eq: S_lin_quad}) corresponds very well with simulations at short times. The deviation at later times occurs due to the build-up of correlations in the system, neglected in the mean field theory. The system size used here is $N_{\text{sys}}=2048$. The parameter values used in {\bf (a)} for ($a,b,b'$) are respectively: {\bf 1} ($0.4, 0.3, 0.3$), {\bf 2} ($0.4, 0.4, 0$), {\bf 3} ($0.4, 0.4, -0.2$), {\bf 4} ($0.4, 0.2, -0.2$), {\bf 5} ($0.4, -0.1, -0.1$), {\bf 6} ($0.4, -0.2, -0.2$), {\bf 7} ($0.4, -0.3, -0.3$), {\bf 8} ($0.4, -0.4, -0.4$), {\bf 9} ($0.2, -0.3, -0.3$) and {\bf 10} ($0.2, -0.4, -0.4$).
}
\label{fig:MeanFieldFits}
\end{figure*}

Within the mean field approximation, the local cross-correlation function
can be expressed in the factorised form $\mathcal{S}(t)={ \frac{1}{N_{\text{sys}}}}\sum_{j=1}^{N_{\text{sys}}}\frac{1}{2}\bigl\langle \delta\sigma_{{\scriptscriptstyle j+1}}-\delta\sigma_{{\scriptscriptstyle j}}\bigr\rangle \bigl\langle \delta\tau_{{\scriptscriptstyle j+\frac{1}{2}}}\bigr\rangle $.
Keeping terms to linear order, we may write $\mathcal{S}(t)$ in terms of the
Fourier variables $\widehat{\delta\sigma}_{k}$ and $\widehat{\delta\tau}_{k}$, we have
\begin{eqnarray}
\mathcal{S}(t) & = & \frac{1}{2N_{\text{sys}}^{3}}
\sum_{j=1}^{N_{\text{sys}}}
\sum_{k=-\pi}^{\pi}\sum_{k'=-\pi}^{\pi}e^{-ikj}e^{-ik'j} \nonumber \\
 &  & \times \left\{ e^{-ik}-1\right\} e^{-ik'/2}\;\left\langle \widehat{\delta\sigma}_{k}(t)\;\widehat{\delta\tau}_{k'}(t)\right\rangle.
\label{eq: S_in_sigma_tau}
\end{eqnarray}
We further write the time dependent correlator $\left\langle \widehat{\delta\sigma}_{{\scriptscriptstyle k}}(t)\;\widehat{\delta\tau}_{{\scriptscriptstyle k'}}(t)\right\rangle $
in terms of the height fields $\widehat{h}_{k}^{P,Q}(t)$. Noting
that the terms $\widehat{h}_{k}^{P}\widehat{h}_{-k}^{P}$ and $\widehat{h}_{k}^{Q}\widehat{h}_{-k}^{Q}$
do not contribute to $\mathcal{S} (t)$ since they are even functions of $k$,
we arrive at the following expression
\begin{eqnarray}
\mathcal{S}(t) & = & \frac{\mathcal{C}}{8N_{\text{sys}}^{2}}\sum_{k=-\pi}^{\pi}\,\left(e^{ik/2}-e^{-ik/2}\right)^{3}\nonumber \\
 &  & \times \left\langle \widehat{h}_{k}^{P}(t)\widehat{h}_{-k}^{Q}(t)-\widehat{h}_{k}^{Q}(t)\widehat{h}_{-k}^{P}(t)\right\rangle .\label{eq: S_hpq}
\end{eqnarray}

Expressing $\widehat{h}_{k}^{P,Q}(t)$ in terms of the initial
condition and noise evolutions through Eq.~(\ref{eq: heqs}), we see
that $\mathcal{S}(t)$ can be written as the sum of two terms $\mathcal{S}_{1}(t)\,+\,\mathcal{S}_{2}(t)$. A detailed derivation of these expressions is presented in Appendix \ref{App: Derivation-of-integrals}. Each of the two terms has a distinct physical origin.
$\mathcal{S}_{1}$ [Eq.~(\ref{eq: S1})] involves the random initial configuration
through the correlator (i) $\left\langle \widehat{h}_{{\scriptscriptstyle k}}^{P}(0)\widehat{h}_{{\scriptscriptstyle -k}}^{Q}(0)\right\rangle $,
while $\mathcal{S}_{2}$ [Eq.~(\ref{eq: S2})] involves the noise through the
correlator (ii) $\left\langle \widehat{\eta}_{{\scriptscriptstyle k}}^{P}(t')\widehat{\eta}_{{\scriptscriptstyle -k}}^{Q}(t'')\right\rangle $. In the continuum limit, we find that $\mathcal{S}(t)$ is a linear combination of two integrals $\mathcal{I}_{1}(t)$
and $\mathcal{I}_{2}(t)$, derived from $\mathcal{S}_{1}$ and $\mathcal{S}_{2}$ respectively. Their forms are presented in Eqs.~(\ref{eq: I1}) and (\ref{eq: I2}). The resulting form of $\mathcal{S}(t)$ in Eq.~(\ref{eq: S_M=000026D}) involves
strengths for the random initial state, $M^{\sigma}=1-\sigma_{0}^{2}$
and $M^{\tau}=1$. We further consider the noises $\eta^{\sigma}$
and $\eta^{\tau}$ to be similarly distributed, implying that their
strengths can be related through $D^{\sigma}/\left(1-\sigma_{0}^{2}\right)=D^{\tau}=D$; arising from non-zero $\sigma_{0}$. From the
definitions of $c$ and $\mathcal{C}$, we have $c/\mathcal{C}=b+b'$ and $c ~\mathcal{C}=2a(1-\sigma_{0}^{2})$.
Therefore we may re-write $\mathcal{S}(t)$ as 
\begin{eqnarray}
\mathcal{S}(t) & = & \frac{\left(1-\sigma_{0}^{2}\right)}{2}\left(2a+b+b'\right)\left[\mathcal{I}_{1}(t)+D\:\mathcal{I}_{2}(t)\right].
\label{eq: S_final}
\end{eqnarray}
The parameter $D$ determines the relative contributions
of the two integrals to $\mathcal{S}(t)$. The integral $\mathcal{I}_{1}(t)$
is linear to the leading order in $t$, whereas integral $\mathcal{I}_{2}(t)$
is quadratic (refer to Appendix \ref{App: Derivation-of-integrals}). To second order, the function $\mathcal{S}(t)$ is given by
\begin{eqnarray}
\mathcal{S}(t) & =\frac{\left(1-\sigma_{0}^{2}\right)}{2}\left(2a+b+b'\right)\biggl(t\;- & \left.3\left[\nu+\frac{D}{2}\right]t^{2}\right)\nonumber \\
 &  & +O(t^{3}).
 \label{eq: S_lin_quad}
\end{eqnarray}
The expression of the linear slope from
mean field theory matches our exact calculation in Section \ref{exact_S_section}. Moreover,
the linear and quadratic terms in Eq.~(\ref{eq: S_lin_quad}) always
have opposite signs, for any $a,\,b,\,\mathrm{and}\;b'$. Hence at
early times, $\mathcal{S}(t)$ is expected to show an extremum, as seen in Fig.~\ref{fig:Different-features-of}, and in agreement with simulations.

\subsection{Comparison with simulations}

The integrals $\mathcal{I}_{1}(t)$ and $\mathcal{I}_{2}(t)$ in Eqs.~(\ref{eq: S1}) and (\ref{eq: S2})
can be evaluated numerically. Using Eq.~(\ref{eq: S_final}), we derive the analytical evolution of $\mathcal{S}(t)$ from the mean field theory. In Fig.~\ref{fig:MeanFieldFits} we have compared the analytical $\mathcal{S}(t)$ with the plots
from simulations. For the value of parameter $D=1$, we observe that
the correspondence holds very well at short times across the entire
phase diagram [Fig.~\ref{fig:MeanFieldFits} (a)], and also for varying mean particle density
$\rho(\bullet) = (1 + \sigma_{0})/2$ [Figs.~\ref{fig:MeanFieldFits} (b) and (c)]. Therefore despite neglecting correlations, the linearized
mean field theory succeeds in describing the early time behavior
of $\mathcal{S}(t)$. This is because in our system we have chosen an initial
state without any correlations, which is exactly of the form assumed
by mean field theory.

The expression in Eq.~(\ref{eq: S_final}) leads to different behaviors
of $\mathcal{S}(t)$ in the ordered and disordered regimes, since the constant
$c$ which enters in the eigenvalues [Eq.~(\ref{eq: eigenvalue})]
can be real, imaginary or zero across the phase boundary. We
discuss the different regimes separately below.

\emph{Disordered regime} ($b+b'<0$): The constant $c$ is imaginary
and its magnitude determines the speed of the density-tilt kinematic
wave. The local cross-correlation $\mathcal{S}(t)$ approaches a constant saturation
value, as seen in simulations. For the case of bunchwise balance ($2a+b+b'=0$),
the state remains totally disordered (uncorrelated), and satisfies
the mean field condition of absence of correlations at all times.
Outside the bunchwise balance plane, and beyond the short time correspondences,
the simulations of $\mathcal{S}(t)$ depart from their respective mean field
analogs at later times, attaining non-zero constant values subsequently
in the steady state. These departures result from the build-up of
correlations between particles, which have been neglected in the mean
field theory. In Fig.~\ref{fig:MeanFieldFits} (a) we see that the departure sets in earlier,
and to a greater extent as we move farther away from bunchwise balance.
This trend indicates an increase of the correlation length as we move
away from the bunchwise balance locus, towards FDPO. This is consistent
with the hypothesis \cite{Barma2019Mixed-Order} that
the correlation length diverges as the transition locus is approached
from the disordered phase, indicating a mixed order transition, as
discussed at the end of Section \ref{sec:The-Light-Heavy-Model}.

\emph{FDPO} ($b+b'=0$): The condition $c=0$ identifies the order-disorder
phase boundary, where the tilt field evolves autonomously and governs
the evolution of the particle field. 
Within the linearized theory,
$\mathcal{S}(t)$ attains a constant saturation value in the thermodynamic
limit, whereas simulations indicate a slow decay $\sim t^{-\phi}$
with $\phi \simeq 0.11$.

\emph{Ordered regime} ($b+b'>0$): $c$ is real, giving rise to an
instability which leads to a divergence of $\mathcal{S}(t)$. In reality, non-linearities
curb the runaway behavior predicted by the linearized theory, and
in fact $\mathcal{S}(t)$ approaches zero in the thermodynamic limit, corresponding
to coarsening towards ordered states as seen in simulations.

\section{Late-Time behavior of $\mathcal{S}$}
\label{section_late_time_S}

In this Section, we discuss the evolution of $\mathcal{S}(t)$ at late times,
particularly during coarsening towards the phase separated steady
states in the ordered regime. Unlike quantities such as two-point
correlation functions, which have been employed routinely to study
out-of-equilibrium systems approaching steady state, $\mathcal{S}(t)$ is a \emph{local} quantity which also captures the extent of coarsening towards ordered phases. We show that the average length of irreducible sequences provides a good estimation of the coarsening length scale at late times.
We also discuss the occurrence of a stretch
of time where $\mathcal{S}(t)$ decays with a diffusive power-law preceding
the onset of coarsening, as predicted by the linearized mean field
theory.

\subsection{Coarsening with two point correlation functions
\label{subsec:Coarsening-with-two-point}}

Earlier studies of coarsening pertaining to the ordered phases in
the LH model have shown that the two-point correlation functions exhibit
scaling, as in phase ordering kinetics \cite{bray2002phase-ordering}. The particle density correlation $G\left(r,t\right)=\left\langle \sigma_{j}(t)\,\sigma_{j+r}(t)\right\rangle $
has the following scaling form:
\begin{eqnarray}
G\left(r,t\right) & = & g\left(\frac{r}{\mathcal{L}(t)}\right).\label{eq: CorrelationFnScaling}
\end{eqnarray}
Here $\mathcal{L}(t)$ represents a coarsening length scale
which grows in time typically, but not always, as $\mathcal{L}(t) \sim t^{1/z}$, where $z$ is the dynamic exponent. The arrangements of particles and tilts situated
within a stretch of length $\mathcal{L}(t)$ resemble those in the
steady state of a finite system of size $N_{\text{sys}}=\mathcal{L}(t)$.
The manner in which $\mathcal{L}(t)$ grows with $t$ depends on the
ordered phase towards which the system coarsens. For instance, while
heading towards the SPS phase, the system undergoes very slow coarsening proceeding through an activation process \cite{LBR2000SPS}, leading to $\mathcal{L}(t)\sim\log t$, verified numerically \cite{shaurithesis2020}. In the case of IPS and FPS however,
the system coarsens faster, as a power-law $\sim t^{1/z}$ with $z \simeq 2$ \cite{Chakraborty2016FastDynamics-LH}. Further, the system
also undergoes coarsening with $z \simeq 1.5$ as it approaches
steady state on the transition line of order and disorder, i.e. FDPO \cite{Das2000fdpoPRL,Das2001fdpoPRE}.

A particularly significant feature of the scaling function
$g(y)$ in Eq.~(\ref{eq: CorrelationFnScaling}) is its behavior
at small argument $y$
\begin{eqnarray}
g\left(y\right) & = & \text{\ensuremath{m_{0}^{2}\left[1-g_{1}\left|y\right|^{\alpha}\right]\,},\ensuremath{\:\:}\ensuremath{\left|y\right|\ll1}}.\label{eq: Correlation_small_y}
\end{eqnarray}
In any ordered phase, the intercept of $g(y)$ as $y\rightarrow0$
is equal to the long-range order, $m_{0}^{2}$  \cite{bray2002phase-ordering}. In the case of the three ordered phases SPS, IPS
and FPS, we have $m_{0}^{2}=1$ \cite{Chakraborty2017LH-statics}, whereas in FDPO, $m_{0}^{2}\simeq0.71$
\cite{Das2001fdpoPRE}. For the ordered states SPS, IPS and FPS, the clusters of \emph{H}
and \emph{L} particles are separated by sharp interfaces. This implies
a linear fall of $g(y)$ for small $y$, i.e. the exponent $\alpha=1$
in Eq.~(\ref{eq: Correlation_small_y}). This is consistent with the
Porod Law \cite{PorodLaw}, observed normally in phase ordering with
a scalar order parameter \cite{bray2002phase-ordering}. However in FDPO, the clusters
are separated by broad interfacial regions, smaller than but of the
order of $\mathcal{L}(t)$. Consequently, the scaling function $g(y)$
displays a cusp singularity as $y\rightarrow0$ with exponent $\alpha<1$
$(\simeq0.15)$ \cite{kapri2016op-scaling}, indicating the breakdown of the Porod Law.

\begin{figure}[t!]
\includegraphics[scale=0.30]{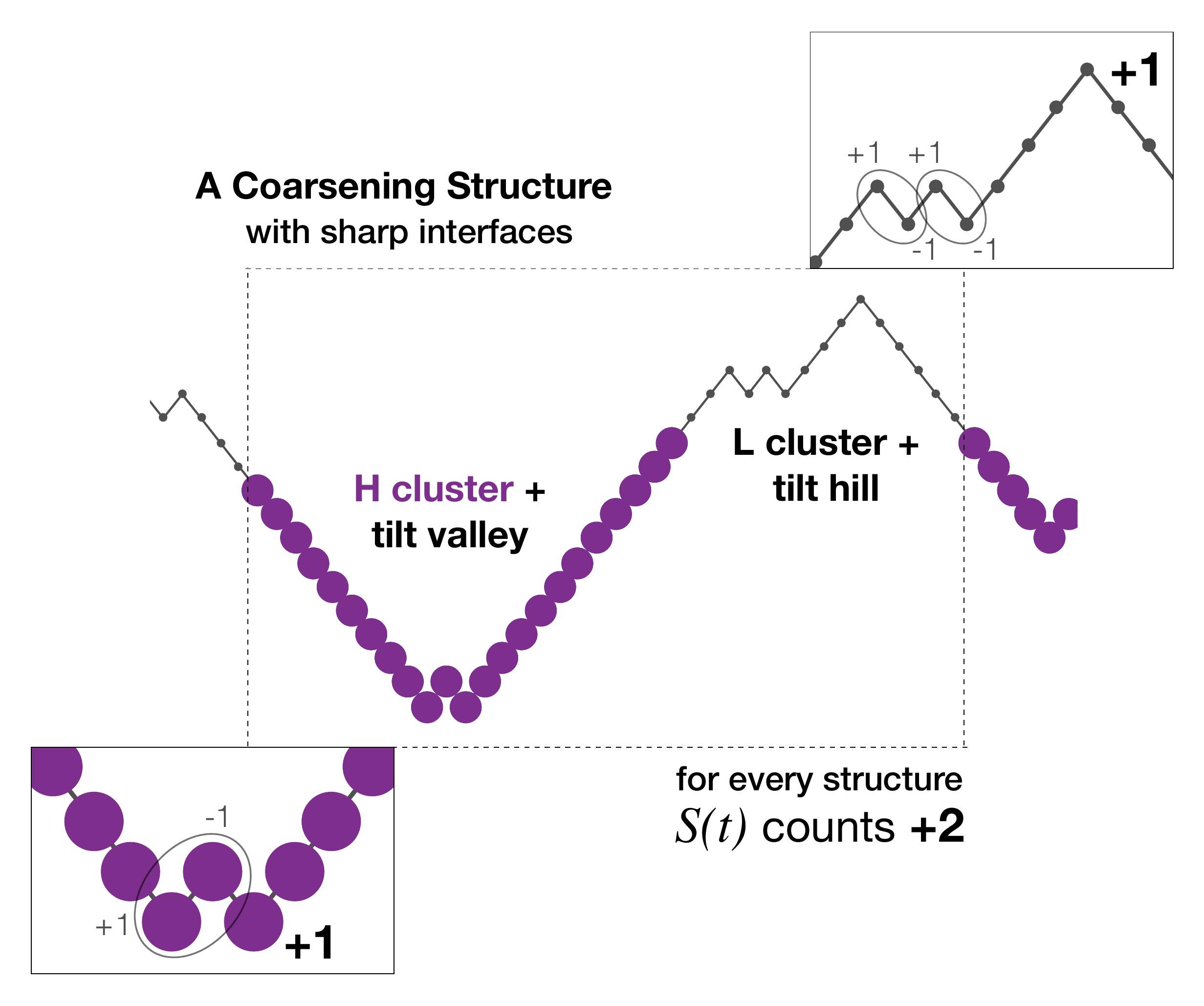}
\caption{\emph{S} as a counter of coarsening `structures' in the LH lattice. The \emph{H} particles are represented by colored circles, while the \emph{L} particles and tilts are represented together by linespoints in grayscale. Every macroscopic valley filled by a cluster of
$H$ particles contains only one triad valley $\diagdown_{\bullet}\diagup$
in excess of the number of residing triad hills $\diagup^{\bullet}\diagdown$,
which is counted as $+1$. All the other triads $\diagdown_{\bullet}\diagup$
($S=+1$) and $\diagup^{\bullet}\diagdown$ ($S=-1$) inside the macroscopic
valley can be grouped into non-contributing pairs. Similarly, every
macroscopic hill overlapping with an L cluster contributes $+1$ to
\emph{S}, by means of its one excess triad $\diagup^{\circ}\diagdown$
($S=+1$). \label{fig: S-Counter}}
\end{figure}

The discussion above is consistent with the following picture: a coarsening
landscape predominantly comprises several large, slowly evolving structures,
whose typical size at time $t$ corresponds to the coarsening length
scale $\mathcal{L}(t)$. Two such adjacent structures $c_{1}$ and
$c_{2}$ undergo sequential mergers over a timescale $t^{*}\sim\mathcal{L}^{z}(t)$,
thus forming a larger structure $c_{1}'$. Assuming that a local
steady state is reached within $\mathcal{L}(t)$ by time $t$, and
referring to the ordered steady state profiles depicted in Fig.~\ref{fig: phase_diagram} (C),
we infer that the coarsening structures typically consist of \emph{H}-particle
clusters overlapping with macroscopic valleys of tilts, along with
their neighboring \emph{L}-particle clusters overlapping with macroscopic
hills. Moreover, these structures also include the interfacial regions
between the\emph{ H}-particle valleys and \emph{L}-particle hills,
which may be quite broad in the case of FDPO.

\subsection{$S(t)$ as a local indicator of coarsening}

Besides its `microscopic' interpretation in Section \ref{S_section}
as a local measure of cross-correlation, $S(t)$ also counts the number
of coarsening structures in the lattice at any time $t$. As discussed
earlier, the sizes of these structures $\mathcal{L}(t)$ may be quite
large.

On the basis of our understanding that a stretch of length $\mathcal{L}(t)$
in an ordered phase is mainly composed of \emph{H}-particle valleys
and\emph{ L}-particle hills, every coarsening structure is expected
to contribute $+2$ to $S(t)$. This is illustrated in Fig.~\ref{fig: S-Counter},
which shows how compact structures with sharp interfaces in a typical
coarsening landscape are counted by $S(t)$. Within every macroscopic
valley filled by a cluster of \emph{H} particles at any given time,
all local triad valleys $\diagdown_{\bullet}\diagup$ and hills $\diagup^{\bullet}\diagdown$
can be grouped into pairs whose contributions to $S(t)$ cancel, except
for\emph{ a} \emph{single} remaining triad valley $\diagdown_{\bullet}\diagup$
which contributes $+1$. Likewise, every macroscopic hill overlapping
with a cluster of \emph{L} particles has \emph{a} \emph{single} local
hill $\diagup^{\circ}\diagdown$ in excess, which also contributes
$+1$. In Section \ref{Coarsening_results} we show that in the ordered phases, the coarsening length scale $\mathcal{L}(t)$ can be extracted from the late time behavior of the disorder averaged correlation $\mathcal{S}(t)$ through the relation
\begin{eqnarray}
\mathcal{S}(t) & \sim & \frac{1}{\mathcal{L}(t)}.\label{eq: CoarseningLength}
\end{eqnarray}

During coarsening towards FDPO however, the coarsening
structures have broad interfacial regions which also contribute
substantially to $S(t)$. Thus the magnitude of $S(t)$ does not directly reflect the number of coarsening structures. Nevertheless $\mathcal{S}(t)$ shows a slow decay at late times. 

\subsection{Irreducible sequences and coarsening length scale}

\begin{figure}[t!]
\includegraphics[scale=0.27]{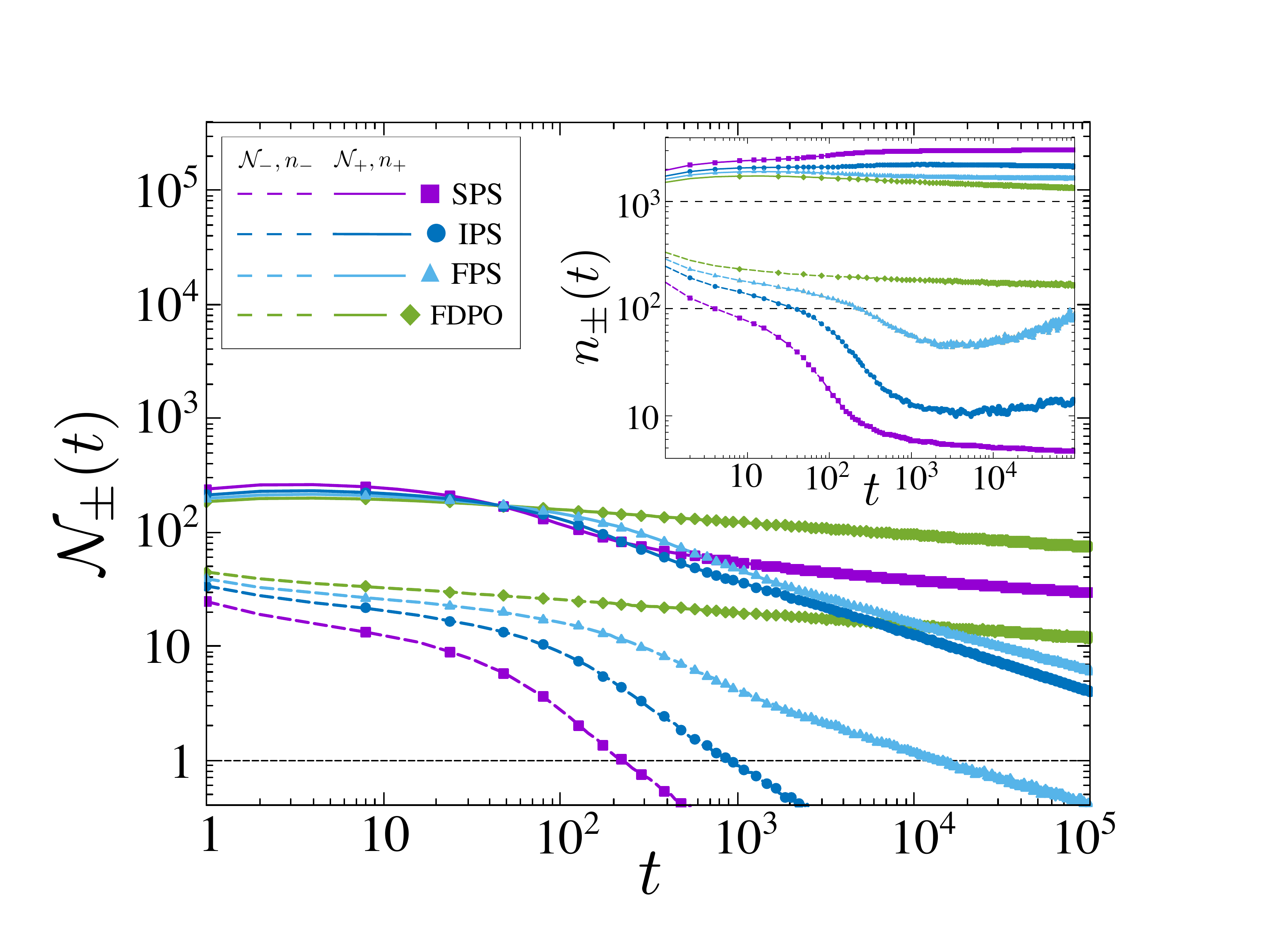}
\caption{A log-log plot of $\mathcal{N}_{\pm}(t)$, the average number of irreducible sequences of types $+$ and $-$ [defined in Eq.~(\ref{plus_minus_definitions})] with time $t$ during coarsening towards the ordered phases, SPS, IPS, FPS and FDPO. The parameter values chosen here are the same as in plots {\bf 1}-{\bf 4} in Fig.~\ref{fig:Early-time-evolution}.
In the regime of the parameter space that we study $(a >0,b+b'\ge0)$, the $+$ sequences are favored, whereas the $-$ sequences are unfavored and their number tends to zero at late times. {\bf Inset}: Log-log plot of the time evolution of the average number of lattice sites $n_{\pm}(t)$ within the $+$ and $-$ sequences, for different ordered phases. The system size used here is $N_{\text{sys}} = 2048$.
\label{fig: N_plus}}
\end{figure}

Next, we show that the coarsening length scale discussed above corresponds to the length scale of the irreducible sequences involving interfaces and bends described in Section \ref{bunchwise_section}. 
The dynamics of the system proceeds through local updates of the particles and tilts, or equivalently, the interfaces and bends. Since we have established a direct relation between the numbers of these sequences and the local cross-correlation in Eq.~(\ref{eq_S_irreducible_connection}), this naturally leads to a length scale describing $S$, namely the lengths of the irreducible sequences. This can be established as follows, at any given time, 
the sites of the system can be grouped into sequences that are reducible, irreducible, as well as sites not belonging to any sequence. 
Focusing specifically on the irreducible sequences which govern $S$, we may then assign spin variables to all the sites of the system with $+$ for sequences of type  $(\langle)\rangle$, $-$ for sequences of type  $\langle(\rangle)$, and $0$ for sites belonging to reducible sequences as well as sites not belonging to any sequence. These $\pm 1$ indices are assigned to all sites from the start to end of an irreducible sequence.
At any time $t$, there are a finite number of sites in each of these states given by $n_{+}(t)$, $n_{-}(t)$ and $n_{0}(t)$ respectively, with $n_{+}(t) + n_{-}(t)+ n_{0}(t) = 2 N_{\text{sys}}$. 
Additionally, we assume that the size of the irreducible sequences are well-described by their {\it average} length $\mathcal{L}_{+}$ and $\mathcal{L}_{-}$, such that the total number of sequences are given by $N_{+} =  n_{+} /\mathcal{L}_{+}$ and $N_{-} =   n_{-}/\mathcal{L}_{-}$. We can then compute $S$ for each configuration, using Eq.~(\ref{eq_S_irreducible_connection}), as $S(t) = (2 n_{+}/\mathcal{L}_{+}) - (2 n_{-}/\mathcal{L}_{-})$. As discussed in Section \ref{bunchwise_section}, one type of sequence is `favored' whereas the other is `unfavored', i.e. the system preferentially evolves towards favored structures.
In our simulations we study the regime of the parameter space $a >0,b+b'\ge0$, and therefore $R = 2 a + b + b' > 0$. Thus irreducible sequences of type $+$ dominate at late times. This behavior is illustrated in Fig.~\ref{fig: N_plus}, where we plot $\mathcal{N}_{\pm}$, the number of $\pm$ sequences averaged over different evolutions, showing the asymmetry in the number of $+$ and $-$ sequences at late times.

\begin{figure}[t!]
\includegraphics[scale=0.274]{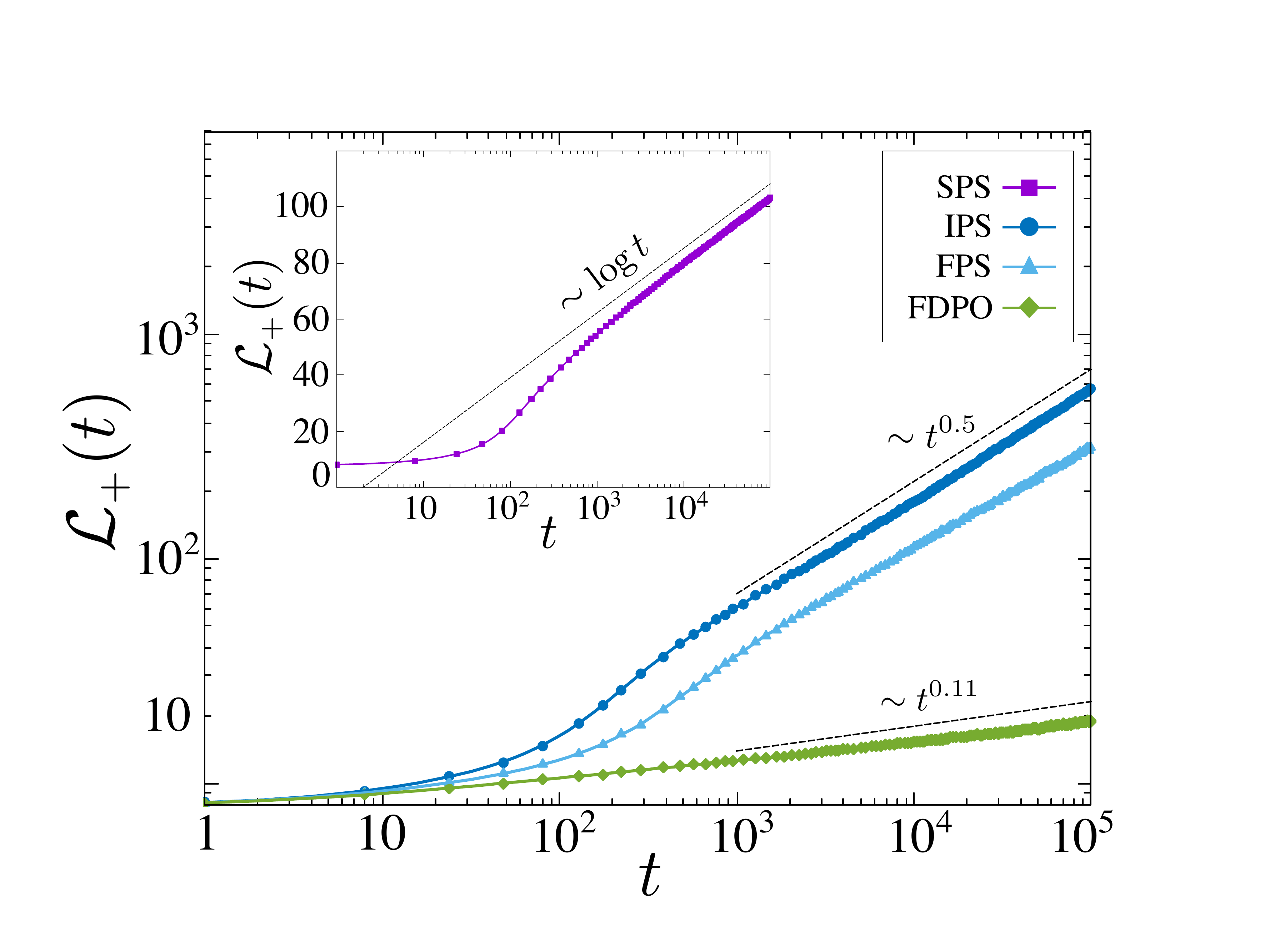}
\caption{Growth of $\mathcal{L}_{+}(t)$, the average length of  irreducible sequences of type $+$ with time $t$ as the system coarsens towards the ordered phases. A log-log plot of $\mathcal{L}_{+}(t)$ in the different ordered phases. IPS, FPS and FDPO display a power law growth in the length of sequences.
{\bf Inset}: Linear-log plot of $\mathcal{L}_{+}(t)$ in the SPS phase, displaying a logarithmic increase. The parameter values chosen here are the same as in plots {\bf 1}-{\bf 4} in Fig.~\ref{fig:Early-time-evolution}.
The system size used here is $N_{\text{sys}} = 2048$.
\label{fig: L_plus}}
\end{figure}

Next, we make the assumption that the numbers of sites $n_{+}$ in the favored sequences remains constant, or near constant over the coarsening dynamics, indicating that the favored sequences primarily lengthen through mergers. This behavior is illustrated in the inset of Fig.~\ref{fig: N_plus}.
This leads us to an estimate of the scaling of the local cross-correlation $\mathcal{S} \sim 1/\mathcal{L}_{+}$. In Fig.~\ref{fig: L_plus} we show the evolution of the average length of the $+$ sequences for the various phases in the LH model. We find that indeed the length scale associated with the irreducible sequences is governed by the same coarsening exponents as the local cross-correlation function $\mathcal{S}$. We note that the length scale of irreducible sequences in fact provides a coarsening length scale that can be measured in {\it every} configuration and not only in the average over evolutions.

Finally, we note the direct relationship between the coarsening structures introduced in the previous subsection and the irreducible sequences. 
A coarsening structure in the ordered regime consists of a cluster of heavy particles within a valley adjacent to a cluster of light particles on a hill, as can be seen in Fig.~\ref{fig: S-Counter}. This is naturally described in terms of interfaces and bends defined in Section \ref{S_section} as $( ... \langle ... ) ... \rangle$, which is an irreducible sequence of type $+$, as given in Eq.~(\ref{plus_minus_definitions}).
Therefore the length of irreducible sequences provides a direct measure of the length of the coarsening structures, through which the coarsening length scale can be probed.\\

\subsection{Coarsening results from $\mathcal{S}(t)$}
\label{Coarsening_results}

\begin{figure*}[t!]
\includegraphics[scale=0.51]{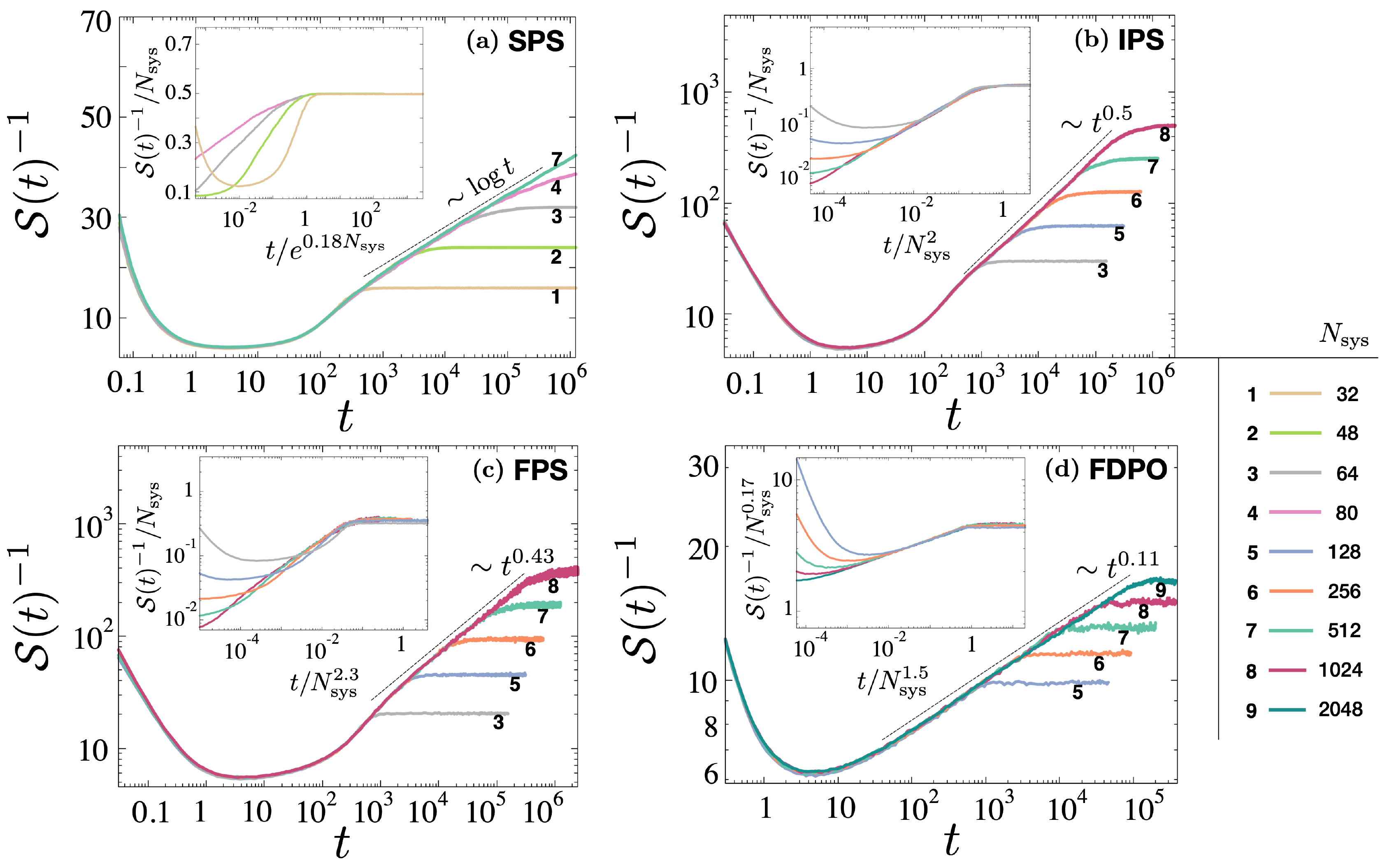}
\caption{Coarsening and finite size effects of $\mathcal{S}(t)$ in the ordered phases {\bf (a)} SPS, {\bf (b)} IPS, {\bf (c)} FPS, and {\bf (d)} the order-disorder separatrix (FDPO). The parameter values chosen here are the same as in plots {\bf 1}-{\bf 4} in Fig.~\ref{fig:Early-time-evolution}.
{\bf (a)} A linear-log plot of $1/\mathcal{S}(t)$ in the SPS phase displays logarithmic coarsening. {\bf Inset:} The saturation value in the steady state scales with $N_{\text{sys}}$ and occurs at a characteristic timescale that diverges exponentially with $N_{\text{sys}}$.
{\bf (b)}-{\bf (d)} The log-log plots of $1/\mathcal{S}(t)$ in the other phases exhibit power law coarsening, with a scaling form consistent with Eq.~(\ref{eq: ScalingFormIPS}) (shown in the {\bf Insets}). 
}
\label{fig:Coarsening-and-finite}
\end{figure*}

We discuss below the decay characteristics of $\mathcal{S}(t)$ during
coarsening, its scaling behavior, and finite size effects.

\emph{Coarsening towards ordered states (SPS, IPS and FPS)}: The decay
profile of $\mathcal{S}(t)$ at late times reflects the dynamics of the coarsening
regime. We argue below that $\mathcal{S}(t)$ accurately counts the diminishing
number of coarsening structures, which is inversely proportional to
their growing sizes $\mathcal{L}(t)$. Identifying the coarsening length scale of the system as the average length of the favored irreducible sequences $\mathcal{L}(t) \equiv \mathcal{L}_{+}(t)$, leads to Eq.~(\ref{eq: CoarseningLength}).
Based on this and the discussion on $\mathcal{L}(t)$
in the previous subsection, we expect $\mathcal{S}(t)$ to decay as $\sim1/\log t$
for SPS, and as a power-law $\sim t^{-\phi}$ with $\phi\simeq0.5$
for IPS and FPS. These behaviors are verified by numerical simulation,
as shown in Fig.~\ref{fig:Coarsening-and-finite}. For a finite-sized system, the steady state is reached when $\mathcal{L}(t)$
becomes as large as the system size $N_{\text{sys}}$, i.e. when $t$ is of the order $t_{s}(N_{\mathrm{sys}})$. Beyond this time $\mathcal{S}(t)$ saturates
to a constant value $\mathcal{S}_{ss}\approx A/N_{\text{sys}}$, as discussed earlier
in Section \ref{S_section}. For the SPS phase, $t_{s}(N_{\mathrm{sys}})\sim e^{\lambda N_{\mathrm{sys}}}$, where $\lambda$ is a constant, whereas for IPS and FPS phases, we have $t_{s}(N_{\mathrm{sys}})\sim N_{\mathrm{sys}}^{z}$. The average number of structures in the steady state is proportional to the number of active triads present, i.e. $ A \sim O(1)$, independent of $N_{\text{sys}}$. 

The late-time behavior of $\mathcal{S}(t)$ from numerical simulations is shown for the ordered phases SPS, IPS and FPS in Fig.~\ref{fig:Coarsening-and-finite} (a)-(c). In SPS [Fig.~\ref{fig:Coarsening-and-finite} (a)], we observe that $\mathcal{S}(t)^{-1}$ grows as $\sim \log t$ for $t < t_s(N_{\text{sys}})$, and saturates when $t > t_s(N_{\text{sys}})$. The inset gives evidence that $t_s \sim e^{\lambda N_{\mathrm{sys}}}$, and $\mathcal{S}_{ss} \sim 1/N_{\text{sys}}$. The late time evolution in the IPS and FPS phases can be collapsed by suitable rescaling [insets of Fig.~\ref{fig:Coarsening-and-finite} (b)-(c)]. A generalized form of $\mathcal{S}(t)$, that applies to IPS, FPS as well as FDPO, is given by the scaling ansatz
\begin{eqnarray}
\mathcal{S}(t) & = & \frac{1}{N_{\text{sys}}^{\mu}}F\left(\frac{\mathcal{L}(t)}{N_{\text{sys}}^{}}\right),
\label{eq: ScalingFormIPS}
\end{eqnarray}
where $\mathcal{L}(t) \sim t^{1/z}$. In the limit $t \ll  t_s(N_{\text{sys}})$, as seen in simulations we have $\mathcal{S}(t)\sim t^{-\phi}$, independent of $N_{\mathrm{sys}}$. Thus for small $y$ we have $F(y) \sim y^{-\mu}$ and $\phi=\mu/z$.  For $t\gg  t_s(N_{\text{sys}})$, the system approaches steady state, and as seen in simulations, we have $\mathcal{S}(t)\sim N_{\mathrm{sys}}^{-\mu}$. Consequently, the scaling function $F(y)$ approaches an $O(1)$ constant as $y \rightarrow \infty$. For the IPS and FPS phases we have $\mu=1$. The data presented in Fig.~\ref{fig:Coarsening-and-finite} (b)-(c) is consistent with the scaling form in Eq.~(\ref{eq: ScalingFormIPS}).

\emph{Fluctuation Dominated Phase Ordering (FDPO)}: In this case,
Eq.~(\ref{eq: CoarseningLength}) does not hold as the coarsening
structures have broad interfacial regions which also contribute
substantially to $\mathcal{S}(t)$. Nevertheless, the scaling form in Eq.~(\ref{eq: ScalingFormIPS}) continues to hold. The numerical results in Fig.~\ref{fig:Coarsening-and-finite} (d) show that $\mathcal{S}(t)$ decays as a slow power law $\sim t^{-\phi}$ with $\phi\simeq0.11$ during coarsening, as the system evolves towards a steady state with large fluctuations in the extent of ordering. As discussed in Section \ref{S_section}, the number of structures in steady state for a finite system scales as $A\sim N_{\text{sys}}^{1-\mu}$, consistent with $\mathcal{S}_{ss} \sim N_{\mathrm{sys}}^{-\mu}$. The scaling relation $\phi=\mu/z$ is satisfied with $\mu\simeq0.17$, $\phi\simeq0.11$ and $z\simeq1.5$.

\begin{figure}[t!]
\includegraphics[scale=0.28]{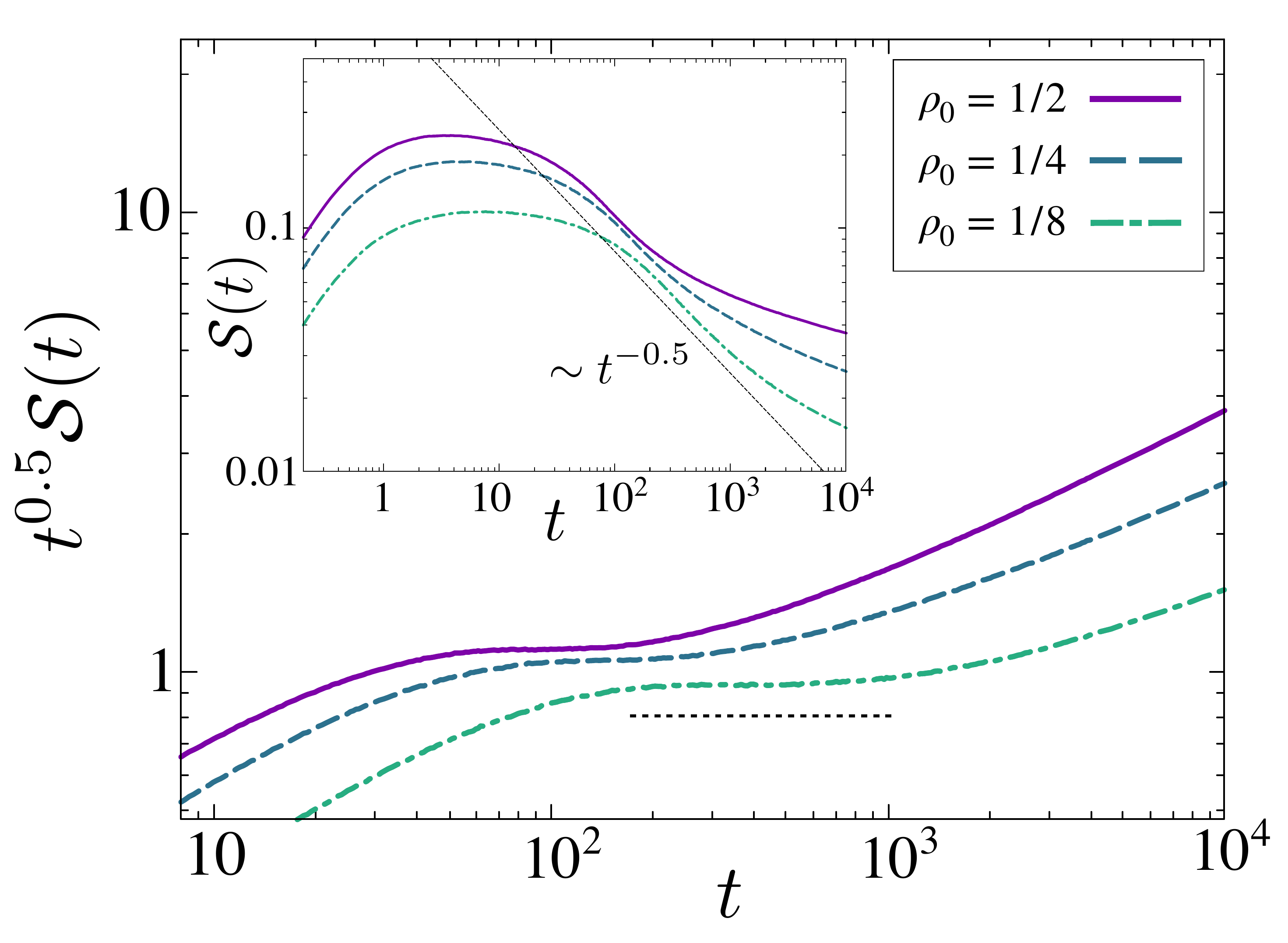}

\caption{Pre-coarsening stretches in the SPS phase for varying mean particle densities $\rho(\bullet)=\frac{1+\sigma_0}{2}=1/2,\,1/4$, and $1/8$, with fixed parameters $a=0.4$ and $b=b'=0.2$. A plot of $t^{0.5}\mathcal{S}(t)$ vs $t$ shows that the time span of the $\sim t^{-0.5}$ pre-coarsening decay increases with decreasing density $\rho (\bullet)$, and can extend to about a decade in time for $\rho (\bullet) =1/8$. {\bf Inset}: $\mathcal{S}(t)$ vs $t$. The system size used here is $N_{\text{sys}}=512$. }
\label{fig:Pre-coarsening-stretch-in}
\end{figure}

In simulations towards the SPS phase, we also observe a power-law stretch over time where $\mathcal{S}(t)$ decays as $\sim t^{-0.5}$, before the onset of the $\sim 1/\log t$ coarsening. The time span of this `pre-coarsening' stretch grows as we lower the particle density $\rho(\bullet) = (1 + \sigma_{0})/2$ in our simulations, and can extend across a decade as shown in Fig.~\ref{fig:Pre-coarsening-stretch-in}. From the expressions of integrals $\mathcal{I}_1(t)$ and $\mathcal{I}_2(t)$ [Eqs.~(\ref{eq: I1}) and (\ref{eq: I2})], we observe that the mean field evolution of $\mathcal{S}(t)$ is governed by the interplay of three timescales (1) $t_{\nu}\sim 1/\nu$ (2) $t_{c}\sim 1/c$  and (3) $t_{\mathrm{ins}}\sim \nu/c^{2}$. The effect of diffusion dominates between time scales $t_{\nu}$  and $t_{c}$, resulting in a diffusive  $\sim t^{-0.5}$ decay of $\mathcal{S}(t)$. The time scale $t_{\mathrm{ins}}$ represents the characteristic time beyond which the linear instability prevails over diffusion. For strong diffusion ($\nu\gg1$) we have $t_{\nu}<t_{c}<t_{\mathrm{ins}}$. Therefore the power-law $\sim t^{-0.5}$ stretch observed in simulations can also exist within the mean field theory. Distinguishing the pre-coarsening effect is difficult in the IPS and FPS phases, as $\mathcal{S}(t)$ decays during coarsening as $\sim t^{-\phi}$, with $\phi$ close to the pre-coarsening exponent $0.5$.

\section{Conclusions and Discussion}
\label{conclusions_section}

In this paper we have studied the LH model, which describes light and heavy particles advecting and interacting with a fluctuating surface. We introduced three new theoretical ideas in this work: bunchwise balance, irreducible sequences and a local cross-correlation function. We established a new condition $2a+b+b' = 0$ in the parameter space of this model, under which the steady state of the system is characterized by an equiprobable measure over all configurations. Furthermore, we showed that this condition is necessary and sufficient for a product measure steady state. This occurs via a novel mechanism which we termed `bunchwise balance', in which the incoming probability current into every configuration from a group of configurations is exactly balanced by the outgoing current to another uniquely specified group of configurations. Next, we identified a local cross-correlation function $\mathcal{S}$, involving the particle density at a site, and its adjacent tilts. We showed that $\mathcal{S}$ is able to capture and distinguish between the properties of different phases that occur in this model. We showed using an exact argument that the initial evolution of $\mathcal{S}(t)$ starting from a totally disordered configuration is linear, with a slope proportional to $R =2a+b+b'$. We then used a set of linearized equations derived from a mean-field expression for the current to describe the early-time dynamics in the LH model, up to quadratic order in time. We provided evidence that the point at which the early time evolution saturates is related to the discreteness of the underlying lattice. We also provided numerical evidence that this mean field theory is able to capture several non-trivial aspects of the evolution of $\mathcal{S}$. Finally we studied the late-time coarsening behavior of the system through the local cross-correlation function, and showed that surprisingly, this local quantity is able to characterize several non-trivial coarsening properties of the system. We also provided numerical evidence that the length of irreducible sequences, which have a direct relation to $\mathcal{S}$, provide an accurate estimation of the coarsening length scale associated with the LH model.

Several interesting directions remain open. As we have shown, the LH model displays an equiprobable steady state through a bunchwise balance mechanism, where the bunches in this model consist of two incoming and two outgoing transitions. It would be interesting to find models that display such a condition with larger bunches, or even unequal numbers of transitions in each bunch. The local cross-correlation function studied in this paper is able to capture the non-trivial coarsening properties in the system which are usually probed through non-local quantities. Indeed, we have established a relationship between the local correlations and the non-local `irreducible' sequences. However, the exact dynamics of these sequences and how these lead to the various non-trivial exponents associated with the coarsening dynamics of the LH model remain to be established. It would be interesting to extend our study of the local cross-correlation $S$ and the concept of irreducible sequences to other multi-species models with a larger number of species. As many of the concepts introduced in this work rely on the one-dimensional nature of the system, it would also be useful to search for generalizations in higher dimensions. Finally, it would be intriguing to use the irreducible sequences introduced in this work to provide a quantitative insight into the nature of typical FDPO configurations, and in particular the structure of the interfacial regions in this regime.


\section*{Acknowledgements}

We acknowledge useful discussions with S.~Chakraborty, S.~Chatterjee, A.~S.~Rajput, \mbox{V.~V.~Krishnan} and R.~Dandekar. A part of the work presented in Sections \ref{S_section} and \ref{linearized_mean_field_section} is discussed in the thesis submitted by S.M. for the degree of Integrated Master of Science (2018) awarded by the UM-DAE Centre for Excellence in Basic Sciences, Mumbai. S.M. would like to thank CEBS Mumbai and TIFR Hyderabad for hospitality and academic support. This project was funded by intramural funds at TIFR Hyderabad from the Department of Atomic Energy (DAE), Government of India. M.B. acknowledges support under the DAE Homi Bhabha Chair Professorship of the Department of Atomic Energy.

\appendix

\section{Derivation of integrals $\mathcal{I}_{1}(t)$ and $\mathcal{I}_{2}(t)$}
\label{App: Derivation-of-integrals}

\begin{figure}[t!]
\includegraphics[scale=0.28]{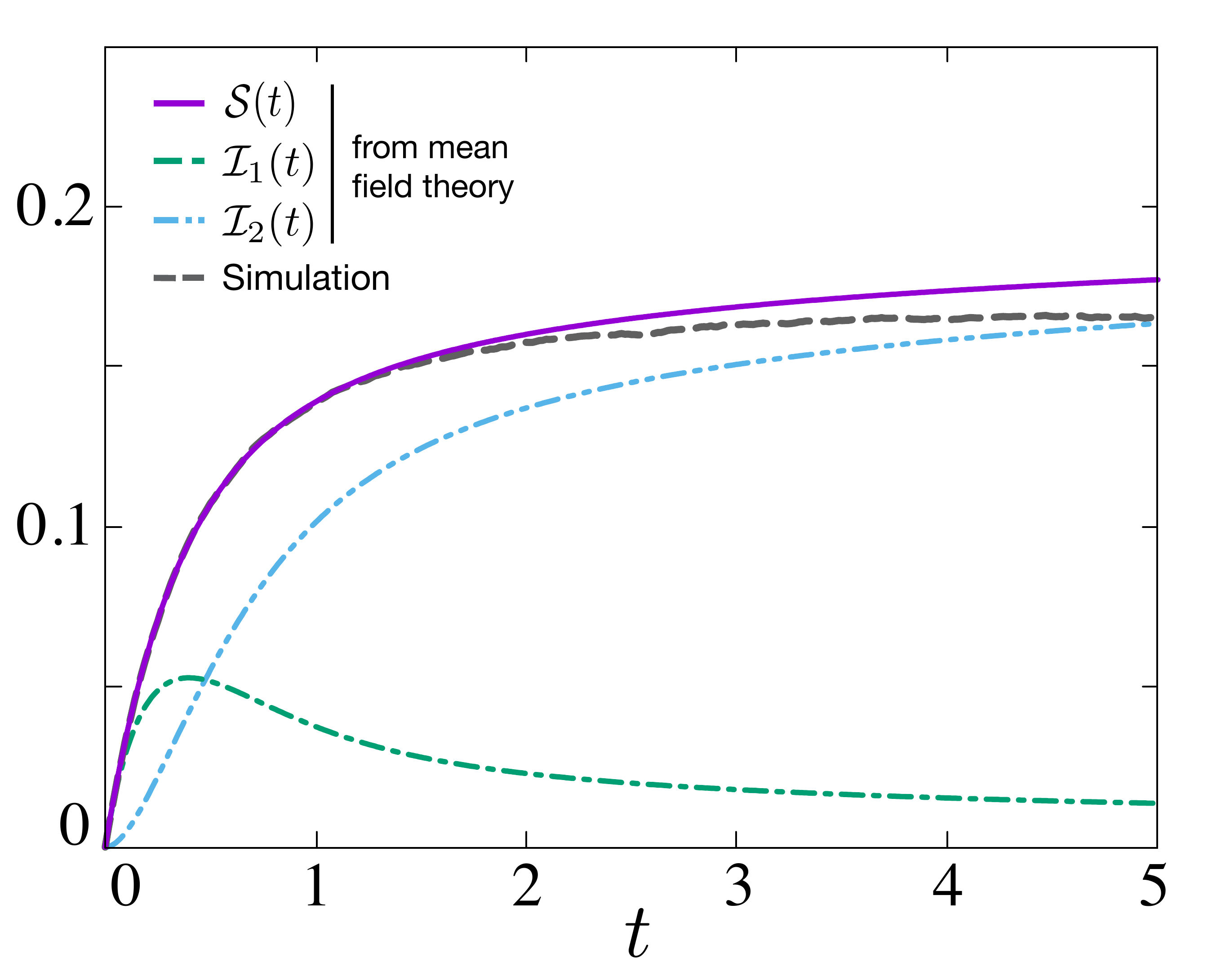}\caption{A typical plot of $\mathcal{S}(t)$ from the linearized mean field theory, shown in
purple. The system shown here evolves from a totally disordered initial state towards an ordered phase (SPS). The two constitutent integrals $\mathcal{I}_{1}(t)$ (in green) and $\mathcal{I}_{2}(t)$ (in blue) encode the effects of the initial state and noise respectively. At short times, the mean field evolution corresponds well with the simulation of $\mathcal{S}(t)$ (in dashed grey). The system size used here is $N_{\text{sys}}=2048$. The values of the parameters are $a=0.4$ and $b=b'=0.2$, with mean particle density $\rho(\bullet)=\frac{1+\sigma_0}{2}=1/2$.}
\label{fig: S_MFT}
\end{figure}

\noindent Expressing $\widehat{h}_{k}^{P,Q}(t)$ in Eq.~(\ref{eq: S_hpq}) in terms of the initial condition and noise evolutions through Eq.~(\ref{eq: heqs}), we see
that $\mathcal{S}(t)$ can be written as the sum of two terms, each with a distinct
physical origin.
\begin{eqnarray}
\mathcal{S}(t) & = & \mathcal{S}_{1}(t)\,+\,\mathcal{S}_{2}(t)
\end{eqnarray}
where
\begin{widetext}
\begin{eqnarray}
\mathcal{S}_{1}(t)\; & = & \quad\frac{\mathcal{C}}{N_{\text{sys}}^{2}}\:\sum_{k=-\pi}^{\pi}\sin^{3}\left(\frac{k}{2}\right) \left\langle \widehat{h}_{k}^{P}(0)\widehat{h}_{-k}^{Q}(0)\right\rangle \left\{ e^{-\left(\lambda_{k}^{+}+\lambda_{-k}^{-}\right)t}-e^{-\left(\lambda_{-k}^{+}+\lambda_{k}^{-}\right)t}\right\}, \label{eq: S1}\\
\nonumber \\
\mathcal{S}_{2}(t)\; & = & \quad\frac{\mathcal{C}}{N_{\text{sys}}^{2}}\:\sum_{k=-\pi}^{\pi}\sin^{3}\left(\frac{k}{2}\right)\left\{ e^{-\left(\lambda_{k}^{+}+\lambda_{-k}^{-}\right)t}\int_{0}^{t}\int_{0}^{t}dt'dt''\;e^{\left(\lambda_{k}^{+}t'+\lambda_{-k}^{-}t''\right)}\left\langle \widehat{\eta}_{k}^{P}\left(t'\right)\widehat{\eta}_{-k}^{Q}\left(t''\right)\right\rangle \right.\nonumber \\
 &  & \qquad\qquad\qquad\qquad\qquad\quad\left.-\;\;e^{-\left(\lambda_{-k}^{+}+\lambda_{k}^{-}\right)t}\int_{0}^{t}\int_{0}^{t}dt'dt''\;e^{\left(\lambda_{-k}^{+}t'+\lambda_{k}^{-}t''\right)}\left\langle \widehat{\eta}_{-k}^{P}\left(t'\right)\widehat{\eta}_{k}^{Q}\left(t''\right)\right\rangle \right\} .\label{eq: S2}
\end{eqnarray}
We next use Eqs.~(\ref{eq: heightcorrelator}) and (\ref{eq: noisecorrelator})
in Eqs.~(\ref{eq: S1}) and (\ref{eq: S2}), and take the continuum limit $\sum_{k=-\pi}^{\pi}\rightarrow\frac{N_{\text{sys}}}{2\pi}\int_{-\pi}^{\pi}dk$. We find that $\mathcal{S}(t)$
is a linear combination of two integrals $\mathcal{I}_{1}(t)$ and
$\mathcal{I}_{2}(t)$ derived from $\mathcal{S}_{1}$ and $\mathcal{S}_{2}$. We have
\begin{eqnarray}
\mathcal{S}(t) & = & \underbrace{\left(\frac{c}{\mathcal{C}}M^{\sigma}+c~\mathcal{C}\,M^{\tau}\right)\,\mathcal{I}_{1}(t)}_{\mathcal{S}_1}\:\:+\:\:\underbrace{\left(\frac{c}{\mathcal{C}}D^{\sigma}+c~\mathcal{C}\,D^{\tau}\right)\,\mathcal{I}_{2}(t)}_{\mathcal{S}_2},
\label{eq: S_M=000026D}
\end{eqnarray}
where the integrals $\mathcal{I}_{1}(t)$ and $\mathcal{I}_{2}(t)$ are given by
\begin{eqnarray}
\mathcal{I}_{1}(t) & = & \frac{1}{4\pi} \int_{-\pi}^{\pi}dk\:\sin\left(\frac{k}{2}\right)\;e^{-4\nu\sin^{2}\left(\frac{k}{2}\right)\,t}\:\frac{1}{c}\sinh\left(2c\sin\left(\frac{k}{2}\right)t\right),\label{eq: I1}\\
\nonumber \\
\mathcal{I}_{2}(t) & = & \frac{1}{\pi} \int_{-\pi}^{\pi}dk\;\int_{0}^{t}dt'\:\sin^{3}\left(\frac{k}{2}\right)\;e^{-4\nu\sin^{2}\left(\frac{k}{2}\right)\,\left(t-t'\right)}\frac{1}{c}\sinh\left(2c\sin\left(\frac{k}{2}\right)\left(t-t'\right)\right).
\label{eq: I2}
\end{eqnarray}
\end{widetext}

In Eq.~(\ref{eq: S_M=000026D}), the correlation strengths for the
random initial state and noises are related as: $M^{\sigma}=1-\sigma_{0}^{2}$, $M^{\tau}=1$, and $D^{\sigma}/\left(1-\sigma_{0}^{2}\right)=D^{\tau}=D$. 
The integral $\mathcal{I}_{1}(t)$  can be expanded up to second order in $t$
as 
\begin{equation}
\mathcal{I}_{1}(t) \sim t - 3\nu t^2 + O(t^3).
\label{simplified_I1}
\end{equation}
Above we have expanded the expression in Eq.~(\ref{eq: I1}) up to second order in $t$ and used the definite integral $\int_{-\pi}^{\pi}dk\,\sin^{2}\frac{k}{2}=\pi$ in the $O(t)$ term and $\int_{-\pi}^{\pi}dk\,\sin^{4}\frac{k}{2}=3\pi/4$ in the $O(t^2)$ term.
Similarly $\mathcal{I}_{2}(t)$ can be expanded up to second order in $t$ as
\begin{equation}
\mathcal{I}_{2}(t) \sim - \frac{3}{2}D t^2 + O(t^3).
\label{simplified_I2}
\end{equation}
Therefore the linear term in $\mathcal{S}(t)$ arises only from $\mathcal{I}_{1}(t)$, while the quadratic term has contributions from both integrals $\mathcal{I}_{1}(t)$ and $\mathcal{I}_{2}(t)$. This behavior is illustrated in Fig.~\ref{fig: S_MFT}.
From the definitions of $c$ and $\mathcal{C}$, we have $c/\mathcal{C}=b+b'$ and $c~\mathcal{C}=2a(1-\sigma_{0}^{2})$. We may therefore re-write $\mathcal{S}(t)$ using the simplified expressions in Eqs.~(\ref{simplified_I1}) and (\ref{simplified_I2}), leading to Eq.~(\ref{eq: S_lin_quad}) in the main text.

\bibliographystyle{apsrev4-1} 
\bibliography{LH_Model_Bibliography.bib}

\end{document}